\documentstyle[eqsecnum,pre,preprint,aps,epsfig]{revtex}

\tightenlines

\newcommand{\bq}{\begin{equation}}
\newcommand{\eq}{\end{equation}}
\newcommand{\bqa}{\begin{eqnarray}}
\newcommand{\eqa}{\end{eqnarray}}

\begin{document}

\draft
\preprint{PM/00-39}

\title{Top quark production at future lepton colliders\\ in the
asymptotic regime}

\author{M. Beccaria$^{a,b}$, F.M.
Renard$^c$ and C. Verzegnassi$^{d,e}$ \\
\vspace{0.4cm} 
}

\address{
$^a$Dipartimento di Fisica, Universit\`a di 
Lecce \\
Via Arnesano, 73100 Lecce, Italy.\\
\vspace{0.2cm}  
$^b$INFN, Sezione di Lecce\\
Via Arnesano, 73100 Lecce, Italy.\\
\vspace{0.2cm} 
$^c$ Physique
Math\'{e}matique et Th\'{e}orique, UMR 5825\\
Universit\'{e} Montpellier
II,  F-34095 Montpellier Cedex 5.\hspace{2.2cm}\\
\vspace{0.2cm} 
$^d$
Dipartimento di Fisica Teorica, Universit\`a di Trieste, \\
Strada Costiera
 14, Miramare (Trieste) \\
\vspace{0.2cm} 
$^e$ INFN, Sezione di Trieste\\
}

\maketitle

\begin{abstract}

The production of a $t\bar t$ pair from lepton-antilepton
annihilation is considered for values of the center of mass 
energy much larger
than the top mass, typically of the few TeV size. In this regime a
number of simplifications occurs that allows to derive the leading
asymptotic terms of various observables using the same theoretical
description that was used for light quark production. Explicit
examples are shown for the Standard Model and the Minimal
Supersymmetric Standard Model cases.

\end{abstract}


\section{Introduction.}

In a number of recent papers \cite{log,mt,slog}, the
production of lepton-antilepton and quark-antiquark pairs from
lepton-antilepton colliders was considered at the one-loop level, with
special emplasis on the "asymptotic" leading behaviour of various
observables, in the case of "light" (i.e. $u,d,s,c,b$) quarks. This
analysis was performed both for the Standard Model (SM)
 and for the Minimal Supersymmetric Standard Model (MSSM) cases, and the
results are fully illustrated in Refs.\cite{log}, \cite{mt},
\cite{slog}. In particular, it was stressed that the leading asymptotic
behaviour is not provided by the known Renormalization Group (RG)
logarithms alone, but from the overall term which is obtained adding
to the RG linear logarithms those of the so-called "Sudakov-type"
\cite{Sudakov}. The latter ones are both of quadratic and of linear
type in the SM, 
but only of linear type in the SUSY additional
contributions that appear in the extra relevant one-loop diagrams. In
the case of final bottom-antibottom production, it was stressed that
important linear logarithmic contributions that are also proportional
to the squared top mass (and, for SUSY diagrams, also to the squared
bottom mass) cannot be neglected, and their numerical effect was
illustrated in several Figures of Ref.\cite{slog}, where special
emphasis was given to the "asymptotic" energy region between $3$ and
$5$ TeV, which is supposed to be covered by the future CERN CLIC
accelerator \cite{CLIC} and, possibly, by a future muon collider
\cite{muon}.\par
A very useful ingredient that was used in the theoretical analysis of
Refs.\cite{log}, \cite{mt}, \cite{slog} is the possibility of 
exploiting an
approach in which several gauge-invariant combinations of one-loop
quantities (self-energies, vertices and boxes) are "subtracted" at
$Z$-peak. This introduces as theoretical input quantities (widths,
asymmetries) that have been measured with extreme precision at LEP1,
SLC. The reward is that of decreasing systematically the number of
theoretical parameters that appear in theoretical models beyond the SM
and of leaving, as one-loop functions, quantities that are from the
begining gauge-invariant and finite \cite{Zsub}, as illustrated in
previous references \cite{Zsubappl}. In principle, this approach can
only be used for final fermion-antifermion pairs that can be
physically produced at the $Z$ peak. Due to this apparently rigid
criterium, it was not applied until now to the case of final
top-antitop production.\par
The aim of this preliminary paper is that of showing that, if one only
considers the "asymptotic" few TeV regime, it is possible to treat
top-antitop production by the same theoretical approach that was
used to describe the production of bottom-antibottom with the only
formal replacement, as a theoretical input, of the (forbidden) $Z$
decay width into top-antitop with the available $Z$ decay into
charm-anticharm. The only residual theoretical difference is that
asymptotic regime description will be provided by known and calculated
one-loop vertices containing the squared top (and bottom) mass. This
will allow to provide theoretical prediction for several observables of
the process, in the few TeV regime, for both the SM and the MSSM cases,
thus completing the already available treatment given for "light"
fermion production in previous references.\par
Technically speaking, the paper will be organised as follows: Section
II will contain a brief kinematical description of the process and of
its simplifications in the asymptotic regime; in Section III the
relevant one-loop diagrams giving rise to the leading asymptotic
contributions will be given for the separate SM and 
MSSM cases. 
Section IV will contain the numerical predictions for various
observables, and finally a short conclusive discussion is made in
Section V. 
The asymptotic expressions of the relevant quantities
that determine the observables of the process are given in Appendix A;
the definitions of the helicity amplitudes and of the observables
specific to the final $t\bar t$ state can be found in Appendix B. 

\section{General description of \lowercase{$t\bar t$}
 production from \lowercase{$l^+l^-$} annihilation}

In full generality, $t\bar t$ production from $l^+l^-$
annihilation at one-loop differs from light fermion production because
two new structures appear in the theoretical description that are a
consequence of the not negligible top mass. This can be visualized in
two equivalent ways, either by considering a "conventional" formalism
(like that used in the light quark case) or by introducing the helicity
amplitudes \cite{ttform}, that are now experimentally more meaningful as
the final top polarization can be measured. To
understand the origin of the extra structures, it will be sufficient to
consider the theoretical expansion of a one loop vertex like that
represented in Fig.1 or 2, with either a photon or a $Z$ entering the
bubble. In full generality, with CP-conserving interactions
one can associate to that diagram the
quantity:

\bq
\Gamma^X_{\mu}=-e^X[\gamma_{\mu}(g^X_{Vt}-g^X_{At}\gamma^5)+{d^X\over
m_t}(p-p')_{\mu}]
\label{3forms}\eq
\noindent
where $X=\gamma,Z$,  $e^{\gamma}=|e|$, $e^Z={|e|\over2s_Wc_W}$
and $p$, $p'$ represent the outgoing $t$, $\bar t$ momenta;
$g^X_{Vt},~g^X_{At},~d^X$ are $O(\alpha)$ one-loop contributions
which in general are
$q^2=(p+p')^2$ dependent. 
The two new quantities $d^X$ enter because the top mass cannot 
now be  neglected and appear
in the various theoretical expressions at one loop, making the overall
number of independent amplitudes of the process to increase from four
(in massless fermion production) to six. This is because the
\underline{three} independent coefficients of eq.(\ref{3forms}) 
will be combined
with the \underline{two} independent coefficients ($g^X_{Vl},~g^X_{Al}$)
of the initial
(massless) lepton vertex.\par
Starting from this general statement, it is now relatively easy to
provide the expressions that appear at one loop e.g. in the helicity
amplitudes formalism. This procedure, which would be essential for a
general description at "moderate" c.m. energies, will be fully
developed in a dedicated forthcoming paper \cite{ttform}. 
But for the specific
purposes of an "asymptotic" energy description, there will be a welcome
simplification. In fact, it is possible to see immediately from
the structure of the one-loop Feynman diagrams
that,  \underline{in the specific cases} of the SM and
in the MSSM, the
coefficients of the new extra
Lorentz structure $(p-p')^{\mu}$ vanish at large $q^2$ like $1/q^2$, 
while those of the "conventional" Lorentz
structures ($\gamma^{\mu},~\gamma^{\mu}\gamma^5$)
can produce either quadratic or linear logarithms.
Therefore, the leading terms of $t\bar t$ production at "asymptotic"
energies are exactly those that would be computed in a 'conventional"
scheme in which the new "scalar" component of eq.(\ref{3forms}) has been
neglected, and \underline{four} independent gauge-invariant
combinations survive that are, formally, equivalent to those of the
final light quark case.\par
At the one loop level, this means that it will be possible to write the
asymptotic invariant scattering amplitude 
of the process $l^+l^-\to t\bar t$ in the
following way:

\bqa
A^{(1)(V,A~only)}_{lt}(q^2,\theta)&=& j^{\mu(\gamma)}_t[{1\over
q^2}(1-\tilde{F}^{\gamma}_{lt}(q^2,\theta))]
j^{(\gamma)}_{\mu,l}+j^{\mu(Z)}_t[{1\over
q^2-m^2_Z}(1-{\tilde{A}^Z_{lt}(q^2,\theta)\over q^2-m^2_Z})]
j^{(Z)}_{\mu,l}
\nonumber\\
&&-j^{\mu(Z)}_t[{1\over
q^2-m^2_Z}{\tilde{A}^{\gamma Z}_{lt}(q^2,\theta)\over q^2}]
j^{(\gamma)}_{\mu,l}  
-j^{\mu(\gamma)}_t[{1\over
q^2-m^2_Z}{\tilde{A}^{Z\gamma}_{lt}(q^2,\theta)\over q^2}]
j^{(Z)}_{\mu,l}  
\eqa
\noindent
where $j^{\mu(\gamma, Z)}_{l,t}$ are the conventional Lorentz structures
used as a basis for the decomposition of the general amplitudes
\cite{Zsub}:
\bq
j^{(\gamma)}_{\mu,f}=-|e|Q_{f}  \gamma_{\mu}
\label{jg}\eq

\bq
j^{(Z)}_{\mu,f}=
-{|e|\over2s_Wc_W}\gamma_{\mu}(g^0_{V,f}-g^0_{A,f}\gamma^5)
\label{jz}\eq
\noindent
with $g^0_{V,f}=I^3_fv_f$, $g^0_{A,f}=I^3_f$, $v_f=1-4|Q_f|s^2_W$.

The four quantities $\tilde{F}^{\gamma}_{lt}(q^2,\theta)$,
$\tilde{A}^Z_{lt}(q^2,\theta)$, $\tilde{A}^{\gamma Z}_{lt}(q^2,\theta)$
and $\tilde{A}^{Z\gamma}_{lt}(q^2,\theta)$ are the generalized
photon and $Z$ one-loop self-energies, 
defined in \cite{Zsub}, which contain
also vertex and box contributions \cite{DS}. 
Starting from these quantities one
will now generalize for $f=t$
the treatment which was done in \cite{Zsub}
in order to construct the
four "subtracted" gauge-invariant functions 
$\widetilde{\Delta}_{\alpha,lf}(q^2,\theta)$, $R_{lf}(q^2,\theta)$,
$V^{\gamma Z}_{lf}(q^2,\theta)$ and $V^{Z\gamma}_{lf}(q^2,\theta)$.
In fact  for the photon component
$\tilde{\Delta}_{\alpha,lt}(q^2,\theta)$ the subtraction is still
performed at $q^2=0$, leaving as theoretical input $\alpha(0)$ and
one-loop quantities that will depend also on $m^2_t$ (and, in
principle, on $m^2_b$).

\bq
\widetilde{\Delta}_{\alpha,
lt}(q^2,\theta)=\widetilde{F}^{\gamma}_{lt}(0,\theta)-
\widetilde{F}^{\gamma}_{lt}(q^2,\theta)
\label{E1}
\eq
\noindent
with
\bqa
\tilde{F}^{\gamma}_{lt} (q^{2}, \theta) = 
F^{\gamma}_{lt} (q^{2}) -
(\Gamma_{\mu,l}^{(\gamma)} , j_{\mu,l}^{(\gamma)}) -
(\Gamma_{\mu,t}^{(\gamma)} , j_{\mu,t}^{(\gamma)})-
q^2 A^{(Box)}_ {\gamma\gamma, lt} (q^{2}, \theta)
\label{Ftilde}
\eqa
\noindent
where the notation $(\Gamma,j)$ means the projection of 
the contribution to the initial or to the final vertex $\Gamma$ on the 
element $j$ of the basis, see eq.(\ref{jg},\ref{jz}) and  
$A^{(Box)}_ {\gamma\gamma, lt} (q^{2}, \theta)$ is the
projection of the $l^+l^-\to t\bar t$ box contribution on
the element $j^{(\gamma)}_tj^{(\gamma)}_l$ of the basis,\cite{Zsub}.

 To illustrate the treatment of the $3$
remaining quantities, we consider briefly the case of
$\tilde{A}^{(Z)}_{lt}(q^2,\theta)$ 
and write its theoretical expression adding and
subtracting the analogous quantity $\tilde{A}^{(Z)}_{lc}(q^2,\theta)$.
After straightforward manipulations, this will lead to the following
situation:\\
a) The theoretical input which will appear in the "Born" term will be
identical with that of charm-anticharm production, in the sense that it
will contain the partial width of $Z$ into $c\bar c$, exactly like in
Ref.\cite{Zsub} for $f=c$.\\
b) The residual one-loop quantity will be the difference between the
non universal vertices and boxes of top production and the
corresponding quantities of charm production. In the notation of
Ref.\cite{Zsub}, this will correspond to 
the introduction of a "modified"
gauge-invariant $\hat{R}_{lt}$ function 
defined in terms of the generalized
$ZZ$ self-energy as:

\bqa
\hat{R}_{lt}(q^2,\theta)&=&R_{lc}(q^2,\theta)
-(\Gamma^{(Z)}_{\mu,t}(q^2)-\Gamma^{(Z)}_{\mu,c}(q^2),j^{(Z)}_{\mu,t})
\nonumber\\
&&
-(q^2-m^2_Z)[
A^{box}_{ZZ,lt}(q^2,\theta)-A^{box}_{ZZ,lc}(q^2,\theta)]
\eqa
\noindent
with the $c$-quark function
\bqa
R_{lc}(q^2,\theta)= I_{Z,lc}(q^2,\theta)-I_{Z,lc}(M^2_Z,\theta)
\label{Rc}
\eqa
\noindent
involving the generalized $ZZ$ function for the production of a
$c\bar c$ pair

\bqa
I_{Z,lc}(q^2,\theta)= {q^2\over q^2-M^2_Z}
[\widetilde{F}^{Z}_{lc}(q^2,\theta)
-\widetilde{F}^{Z}_{lc}(M^2_Z,\theta)]
\label{Ic}
\eqa

A quite analogous procedure can be used for the interference terms
$V^{\gamma Z}$ and $V^{Z\gamma}$. Without entering the full details,
this leads to the introduction in the theoretical input of
forward-backward asymmetry of the $Z$ decay into charm-anticharm and to
the introduction of a new gauge-invariant function:

\bqa
\hat{V}^{Z\gamma}_{lt}(q^2,\theta)&=&V^{Z\gamma}_{lc}(q^2,\theta)
-(\Gamma^{(Z)}_{\mu,t}(q^2)-\Gamma^{(Z)}_{\mu,c}(q^2),
j^{(\gamma)}_{\mu,t})\nonumber\\
&&
-(q^2-m^2_Z)[
A^{box}_{Z\gamma,lt}(q^2,\theta)-A^{box}_{Z\gamma,lc}(q^2,\theta)]
\label{vzg}\eqa
\noindent
with
\bqa
V^{Z\gamma}_{lc}(q^2,\theta)=
\frac{\widetilde{A}^{Z\gamma}_{lc}(q^2,\theta)}{q^2}-
\frac{\widetilde{A}^{Z\gamma}_{lc}(M^2_Z,\theta)}{M_Z^2}
\label{vzgc}
\eqa

\bqa  
 {\widetilde{A}^{Z\gamma}_{lc} (q^{2}, \theta)\over q^2}
&=& {A^{\gamma Z} (q^{2})\over q^2} - {q^2-M^2_{Z}\over q^2}
(\Gamma_{\mu,l}^{(\gamma)}(q^2) , j_{\mu,l}^{(Z)})\nonumber\\
 &&-(\Gamma_{\mu,c}^{(Z)}(q^2) , j_{\mu,c}^{(\gamma)})-(q^2-M^2_{Z})
A^{(Box)}_ {Z\gamma, lc} (q^{2}, \theta)
\label{Azgc}
\eqa

The other interference term $V^{\gamma Z}$ does not require specific
tricks, as in the case of the photon component
$\tilde{\Delta}_{\alpha,lt}(q^2,\theta)$. Its theoretical expression
can be written in fact as :

\bqa
V^{\gamma Z}_{lt}(q^2,\theta)&=&V^{\gamma Z}_{lc}(q^2,\theta)=
\frac{\widetilde{A}^{\gamma Z}_{lt}(q^2,\theta)}{q^2}-
\frac{\widetilde{A}^{\gamma Z}_{lt}(M^2_Z,\theta)}{M_Z^2}
\label{vgz}
\eqa
\noindent
with
\bqa
\frac{\widetilde{A}^{\gamma Z}_{lt}(q^2,\theta)}{q^2}&=&
{A^{\gamma Z} (q^{2})\over q^2}
 - ({q^2-M^2_{Z}\over q^2})
(\Gamma_{\mu,t}^{(\gamma)}(q^2) , j_{\mu,t}^{(Z)})\nonumber\\
 &&-(\Gamma_{\mu,l}^{(Z)}(q^2) , j_{\mu,l}^{(\gamma)})-(q^2-M^2_{Z})
A^{(Box)}_ {\gamma Z, lt} (q^{2}, \theta)
\label{Agz}\eqa
and in eq.(\ref{Agz}) 
no final (top) vertices or boxes appear at $m^2_Z$.\par

The previous equations that we wrote are quite general. From their
expressions one can now determine in a straightforward way the related
asymptotic behaviours. To obtain the latter ones, it will be sufficient
to add to the "light" $c$ quark functions, already computed in
Refs.\cite{log}, \cite{slog}
the extra non universal terms coming from the difference between top
vertices and charm vertices (the difference between boxes will vanish
asymptotically since the latter ones are not producing massive terms,
as already illustrated in those references). This means that the only
extra quantities to be computed for the specific purposes of this paper
are the asymptotic "Sudakov-type" linear logaritms proportional to the
squared top and bottom masses coming from the final top vertex. Their
expressions will be given in the following Section III.\par

\section{Massive one-loop contributions from final top vertices}

For the specific purposes of this paper we shall only be interested in
those contributions to the process of $t\bar t$ production that come
from final top vertices and are proportional either to the squared top
mass or (in practice, only for the SUSY case) to the squared bottom
mass. These are coming from the diagrams shown in Fig.(1) for the SM
and in Fig.(2) for the extra SUSY component of the MSSM, that we
consider separately in this paper. Using our conventional definitions 
\cite{log} of the one-loop vertex $i\Gamma_{\mu}$, we derive the
components of the leading asymptotic behaviour that are proportional to
the quark masses following the same procedure that was used in
Ref.\cite{slog}. The results are given by the following equations:

\bqa
\Gamma^{\gamma}_{\mu}(\mbox{SM},~ \mbox{massive})&\to&
{e\alpha\over24\pi M^2_Ws^2_W}\ \ln q^2
\{m^2_t[(\gamma_{\mu}P_L)+2(\gamma_{\mu}P_R)]
+m^2_b(\gamma_{\mu}P_L)\}
\label{SMgm}\eqa

\bqa
\Gamma^{Z}_{\mu}(\mbox{SM},~ \mbox{massive})&\to&
{e\alpha\over96\pi M^2_Ws^3_Wc_W}\ \ln q^2
\{(3-4s^2_W)m^2_t(\gamma_{\mu}P_L)-8s^2_Wm^2_t(\gamma_{\mu}P_R)
\nonumber\\
&&
+(3-4s^2_W)m^2_b(\gamma_{\mu}P_L)\}
\label{SMZm}\eqa

\bqa
\Gamma^{\gamma}_{\mu}(\chi,~ \mbox{massive})&\to&
{e\alpha\over24\pi M^2_Ws^2_W}\ \ln q^2
\{m^2_t(1+\cot^2\beta)[(\gamma_{\mu}P_L)+2(\gamma_{\mu}P_R)]\nonumber\\
&&
+m^2_b(1+\tan^2\beta)(\gamma_{\mu}P_L)\}
\label{chigm}\eqa

\bqa
\Gamma^{Z}_{\mu}(\chi,~ \mbox{massive})&\to&
{e\alpha\over96\pi M^2_Ws^3_Wc_W}\ \ln q^2
\{(3-4s^2_W)m^2_t(1+\cot^2\beta)(\gamma_{\mu}P_L)\nonumber\\
&&
-8s^2_Wm^2_t(1+\cot^2\beta)(\gamma_{\mu}P_R)
+(3-4s^2_W)m^2_b(1+\tan^2\beta)(\gamma_{\mu}P_L)\}
\label{chiZm}\eqa

\bqa
\Gamma^{\gamma}_{\mu}(H)&\to&
{e\alpha\over24\pi M^2_Ws^2_W}\ \ln q^2
\{m^2_t(\cot^2\beta)[(\gamma_{\mu}P_L)+2(\gamma_{\mu}P_R)]\nonumber\\
&&
+m^2_b(\tan^2\beta)(\gamma_{\mu}P_L)\}
\label{Hgm}\eqa

\bqa
\Gamma^{Z}_{\mu}(H)&\to&
{e\alpha\over96\pi M^2_Ws^3_Wc_W}\ \ln q^2
\{(3-4s^2_W)m^2_t(\cot^2\beta)(\gamma_{\mu}P_L)\nonumber\\
&&
-8s^2_Wm^2_t(\cot^2\beta)(\gamma_{\mu}P_R)
+(3-4s^2_W)m^2_b(\tan^2\beta)(\gamma_{\mu}P_L)\}
\label{HZm}\eqa

\noindent
where $P_{L,R}=(1\mp\gamma^5)/2$. 
$(H)$ denotes the contribution from the SUSY charged and neutral
Higgses of the MSSM (the contribution from the SM Higgs has been
subtracted) and $(\chi)$ denotes the contribution from charginos and
neutralinos of the model.\par
Eqs.(\ref{SMgm}-\ref{HZm}) are the new results of this paper. 
Note that the total MSSM massive
contributions just correspond to the SM ones with the $m^2_t$ terms
being multiplied by $2(1+\cot^2\beta)$ and the $m^2_b$ terms
by $2(1+\tan^2\beta)$, a rule which had already been observed in
Ref.\cite{slog}:

\bqa
\Gamma^{\gamma}_{\mu}(\mbox{MSSM},~ \mbox{massive})&\to&
{e\alpha\over12\pi M^2_Ws^2_W}\ \ln q^2
\{m^2_t(1+\cot^2\beta)[(\gamma_{\mu}P_L)+2(\gamma_{\mu}P_R)]\nonumber\\
&&
+m^2_b(1+\tan^2\beta)(\gamma_{\mu}P_L)\}
\label{MSSMgm}\eqa

\bqa
\Gamma^{Z}_{\mu}(\mbox{MSSM},~ \mbox{massive})&\to&
{e\alpha\over48\pi M^2_Ws^3_Wc_W}\ \ln q^2
\{(3-4s^2_W)m^2_t(1+\cot^2\beta)(\gamma_{\mu}P_L)\nonumber\\
&&
-8s^2_Wm^2_t(1+\cot^2\beta)(\gamma_{\mu}P_R)
+(3-4s^2_W)m^2_b(1+\tan^2\beta)(\gamma_{\mu}P_L)\}
\label{MSSMZm}\eqa

It should be stressed that, as already remarked in ref.\cite{slog}, 
in the \underline{leading} asymptotic SUSY "Sudakov" logarithms 
all the details of the MSSM, in particular the mixing parameters 
appearing in the chargino and neutralino mass matrices, are washed 
out in the large $q^2$ limit, and are
"reshuffled" in subleading constant terms. The only supersymmetric 
parameter of the model that survives asymptotically is $\tan\beta$. 
This will affect the observables of top-antitop production 
in a potentially interesting way that will 
be fully examined in the final discussion.\\

To obtain the expressions of
the various observables of the process $l^+l^-\to t\bar t$, 
one has to add the above
massive contributions to those coming from the massless quark 
situation (in our case, that corresponding to charm production). 
The overall terms are incorporated into the four gauge-invariant 
quantities $\tilde{\Delta}_{\alpha,lt}$, $\hat{R}_{lt}$, 
$V^{\gamma Z}_{lt}$, $\hat{V}^{Z\gamma}_{lt}$ from which all the
observables at one loop can be built in the asymptotic regime of the
$t\bar t$ production process. We have listed the various contributions
in the Appendix A, in the following order; first, the SM contributions,
universal (RG) terms, non-universal massless terms and non-universal
massive ($m^2_t$ and $m^2_b$ dependent) terms; secondly, the additional
SUSY contributions split into the same three groups.\par
Starting from the formulae of Appendices A and B
it is now possible to derive
the theoretical predictions for the leading asymptotic behaviour of the
observables of the process. These are exhibited in the following final
Section IV.

\section{Asymptotic behaviour of the observables of 
\lowercase{$t\bar t$} production}

For the process $l^+l^-\to t\bar t$ we can consider
two sets of observables.
First, we have computed the one-loop effects for 
the same set previously considered in the case of light
quarks: the integrated $e^+e^-\to t\bar t$ cross section denoted by
$\sigma_t$, the forward backward asymmetry $A_{FB,t}$, the
longitudinal polarization asymmetry $A_{LR,t}$ and its
forward-backward polarization asymmetry \cite{afbpol} $A_t$.
Secondly, we have considered a new set of observables 
constructed from the spin
density matrix of the final top \cite{GR};
the averaged top helicity $H_t$, its forward-backward asymmetry
$H_{t,FB}$, as well as the same two observables
$H^{LR}_t$, $H^{LR}_{t,FB}$ in the case of 
longitudinaly polarized $l^{\pm}$ beams.
 Their definition in terms of the top spin density
matrix as well as the general expression of the helicity amplitudes are
given in Appendix B; more details will be found in \cite{ttform}.
In general there could be 4 other observables related to the transverse
top quark polarization, but for asymptotic energies the transverse
polarization degree vanishes like $m_t/\sqrt{q^2}$.\par 

The results for the asymptotic behaviour of each observables
are given by the following equations with the
various terms grouped in the following order:
first in SM, the RG with the
mass scale $\mu$, followed
by the linear and quadratic Sudakov (W diagrams) terms,
the linear and quadratic Sudakov (Z diagrams) terms and finally
the linear Sudakov term arising from
the quadratic $m^2_t$ contribution; then, in bold face,
the SUSY contributions, first the RG (SUSY) term with the
mass scale $\mu$, then the linear 
Sudakov (SUSY) $m_t-$ and $m_b-$ independent term 
(scaled by the common mass $M$),
the linear Sudakov (SUSY) term arising 
from the quadratic $m^2_t$ 
contribution (scaled by a common mass $M'$) and in curly 
brackets the same term to which the $m_b^2\tan^2\beta$ contribution
is added successively for $\tan\beta=10$ and for $\tan\beta=40$. 
This was done
in order to show precisely the origin of the difference
between the total SM prediction and the total SUSY part.We have chosen
to use for simplicity common mass scales $M$ and $M'$
because of the present ignorance of
the physical masses of the charginos, neutralinos and sfermions
appearing in the triangle diagrams of Fig.2 as well as those of the
charged and neutral Higgses appearing together with the top and the
bottom quark. A change of reference scale is equivalent to
the addition of an asymptotically negligible constant term; see
Ref.\cite{slog} and \cite{ttform} for a discussion of this point. \\

We first consider the four observables constructed from the
differential cross section without measuring the final top quark
polarization. In the following equations the various "subtracted" 
Born terms $O^B$ are defined in terms of the
$Z$-peak inputs as explained in Section II.

\bqa
\sigma_{t}&=&\sigma^{B}_{t}\{1+{\alpha\over4\pi}\{(8.87N-33.16
)\ln{q^2\over\mu^2}+(22.79 \ln{q^2\over M^2_W}-
5.53\ln^2{q^2\over M^2_W})
\nonumber\\
&&
+(3.52\ln{q^2\over M^2_Z}-1.67\ln^2{q^2\over M^2_Z})
-~14.21 \ln{q^2\over m^2_t}
\nonumber\\
&&
+{\bf (4.44\,N+11.09
)\,\ln{q^2\over\mu^2}-10.09\, \ln{q^2\over M^2}
-42.63\{-15.33\}\{-27.48\} \ln{q^2\over M'^2}}\}
\label{sigtt}\eqa

\bq
\sigma^{B}_{t}=0.182~\mbox{pb}/q^2(\mbox{TeV}^2)
\eq

\bqa
A_{FB,t}&=&A^{B}_{FB,t}+{\alpha\over4\pi}\{(0.45N-4.85
)\ln{q^2\over\mu^2}-(1.79 \ln{q^2\over M^2_W}+0.17\ln^2{q^2\over M^2_W})
\nonumber\\
&&
-(1.26\ln{q^2\over M^2_Z}+0.06\ln^2{q^2\over M^2_Z})
+0.61 \ln{q^2\over m^2_t}
\nonumber\\
&&
+{\bf (0.22\,N+1.29
)\,\ln{q^2\over\mu^2}-0.23\, \ln{q^2\over M^2}
+1.83\{0.54\}\{-0.68\} \ln{q^2\over M'^2}}
\}
 \ .
\label{AFBt}\eqa

\bq
A^{B}_{FB,t}=0.607
\eq

\bqa
A_{LR,t}&=&A^{B}_{LR,t}+{\alpha\over4\pi}\{(2.06N-22.43
)\ln{q^2\over\mu^2}+(14.75 \ln{q^2\over M^2_W}
-3.54\ln^2{q^2\over M^2_W})
\nonumber\\
&&
+(0.40\ln{q^2\over M^2_Z}-0.49\ln^2{q^2\over M^2_Z})
+3.79 \ln{q^2\over m^2_t}
\nonumber\\
&&
+{\bf (1.03\,N+5.95
)\,\ln{q^2\over\mu^2}-4.03\, \ln{q^2\over M^2}
+11.36\{3.36\}\{-4.23\} \ln{q^2\over M'^2}}
\},
\label{ALRt}\eqa

\bq
A^{B}_{LR,t}=0.336
\eq

\bqa
A_{t}&=&A^{B}_{t}+{\alpha\over4\pi}\{(1.82N-19.77
)\ln{q^2\over\mu^2}
+(8.68 \ln{q^2\over M^2_W}-2.76\ln^2{q^2\over M^2_W})
\nonumber\\
&&
+(0.09\ln{q^2\over M^2_Z}-0.45\ln^2{q^2\over M^2_Z})
+3.69 \ln{q^2\over m^2_t}
\nonumber\\
&&
+{\bf (0.91\,N+5.25
)\,\ln{q^2\over\mu^2}-3.20\, \ln{q^2\over M^2}
+11.06\{3.27\}\{-4.11\} \ln{q^2\over M'^2}}\},
\label{AFBpt}\eqa

\bq
A^{B}_{t}=0.164
\eq

We have then considered the four other observables constructed from
the final top quark helicity and defined in Appendix B. In the
asymptotic regime (i.e neglecting $m^2_t/q^2$ terms and new Lorentz
structures) we notice that, if there were no Box contributions
introducing extra $\theta$-dependences, 
these four observables would be exactly related
to the four previous ones. This is obvious from the fact that, in
this limit, there are only four independent combinations of
photon and $Z$ coupling which describe the differential cross
section for any top or antitop helicity (denoted $G_1$, $G_2$, $G_4$
and $G_5$ in ref.\cite{book}). 
The relations would be the following ones:

\bq
H^{no~Box}_t\equiv-~{4\over3}~A^{no~Box}_t
\label{RHt}
\eq

\bq
H^{no~Box}_{t,FB}\equiv-~{3\over4}~A^{no~Box}_{LR,t}
\label{RHtFB}
\eq

\bq  
H^{LR,no~Box}_t\equiv-~{4\over3}~A^{no~Box}_{FB,t}
\label{RHtLR}
\eq

\bq
H^{LR,no~Box}_{t,FB}\equiv -~{3\over4}
\label{RHtLRFB}
\eq

At non asymptotic energies relations (\ref{RHtFB}) and (\ref{RHtLRFB}),
contrarily to (\ref{RHt}) and  (\ref{RHtLR}),
should be affected not only by Box effects, but also 
by $m^2_t/q^2$ terms and by contributions from the
new Lorentz structures; more
details will be given in \cite{ttform}.\\

Taking the SM box contributions into account we obtain to the following
results:

\bqa
H_t&=&H^{B}_t+{\alpha\over4\pi}\{(-2.42N+26.36
)\ln{q^2\over\mu^2}
+(-15.90 \ln{q^2\over M^2_W}+3.67\ln^2{q^2\over M^2_W})
\nonumber\\
&&
+(-0.74\ln{q^2\over M^2_Z}+0.59\ln^2{q^2\over M^2_Z})
-4.91\ln{q^2\over m^2_t}
\nonumber\\
&&
+{\bf (-1.21\,N-7.00
)\,\ln{q^2\over\mu^2}+4.27\, \ln{q^2\over M^2}
-14.74\{-4.36\}\{+5.49\} \ln{q^2\over M'^2}}\},
\label{Ht}\eqa

\bq
H^{B}_t=-0.219
\eq

\bqa
H_{t,FB}&=&H^{B}_{t,FB}+{\alpha\over4\pi}\{(-1.55N+16.81
)\ln{q^2\over\mu^2}
+(-7.82 \ln{q^2\over M^2_W}+2.65\ln^2{q^2\over M^2_W})
\nonumber\\
&&
+(0.23\ln{q^2\over M^2_Z}+0.37\ln^2{q^2\over M^2_Z})
-2.84\ln{q^2\over m^2_t}
\nonumber\\
&&
+{\bf (-0.77\,N-4.46
)\,\ln{q^2\over\mu^2}+3.02\, \ln{q^2\over M^2}
-8.52\{-2.52\}\{+3.17\} \ln{q^2\over M'^2}}\},
\label{Hfbt}\eqa

\bq
H^{B}_{t,FB}=-0.252
\eq

\bqa
H^{LR}_t&=&H^{LR,B}_t+{\alpha\over4\pi}\{(-0.59N+6.46
)\ln{q^2\over\mu^2}
+(-1.91 \ln{q^2\over M^2_W}+0.23\ln^2{q^2\over M^2_W})
\nonumber\\
&&
+(0.70\ln{q^2\over M^2_Z}+0.08\ln^2{q^2\over M^2_Z})
-0.81\ln{q^2\over m^2_t}
\nonumber\\
&&
+{\bf (-0.30\,N-1.71
)\,\ln{q^2\over\mu^2}+0.31\, \ln{q^2\over M^2}
-2.44\{-0.72\}\{+0.91\} \ln{q^2\over M'^2}}\},
\label{HLRt}\eqa

\bq
H^{LR,B}_t=-0.809
\eq

\bqa
H^{LR}_{t,FB}&=&H^{LR,B}_{t,FB}+{\alpha\over4\pi}\{(0.)
\ln{q^2\over\mu^2}
+(3.24 \ln{q^2\over M^2_W}+(0.)\ln^2{q^2\over M^2_W})
\nonumber\\
&&
+(0.50\ln{q^2\over M^2_Z}+(0.)\ln^2{q^2\over M^2_Z})
+(0.)\ln{q^2\over m^2_t}
\nonumber\\
&&
+{\bf (0.)\,\ln{q^2\over\mu^2}+(0.)\, \ln{q^2\over M^2}
+(0.) \ln{q^2\over M'^2}}\},
\label{HLRFBt}\eqa

\bq
H^{LR,B}_{t,FB}=-0.750
\label{NHLRFBt}\eq

One can check that, indeed apart from the two SM coefficients
of $\ln{q^2\over M^2_W}$ and $\ln{q^2\over M^2_Z}$,
affected by the Box contributions, all other coefficients,
as well as the "Born" terms
satisfy the relations (\ref{RHt})-(\ref{RHtLRFB}). In particular
eq,(\ref{RHtLRFB}) is responsible for the appearence 
of the various zeros in 
eq.(\ref{HLRFBt}). So the physical content of the four observables
constructed with the top helicity is almost the same as that
of the polarized differential cross section. The importance of the
angular dependence in the SM box contribution can be appreciated
from the size of the non zero coefficients in
eq.(\ref{HLRFBt}) and will be illustrated in a figure given below.
A comparison with experimental data on top quark polarization
should be useful for a confirmation of the results obtained from the
cross section and the asymmetries and should constitute a check of the
model (SM or MSSM) and of the absence of unexpectedly large
asymptotic contributions.  
\\

Eqs.(\ref{sigtt})-(\ref{NHLRFBt})
are the main result of this paper. 
To better appreciate their 
message, we have
plotted in the following Figs.(3)-(6) and (7)-(10),
the asymptotic terms, 
with the following convention:
for the cross section, we show the relative effect; for asymmetries and
helicities, 
the absolute effect. To fix a
scale, we also write in the Figure captions 
the value of the (asymptotic) 
"Born " terms and we have put $\mu=M_Z$ for the RG terms and
$M=M'=m_t$ for the SUSY terms. The plots have been
drawn in an energy region between one and ten TeV.\\

 We have
plotted the overall SM value, the overall MSSM value for purposes of
comparison, and the approximate expressions that would be obtained by
only retaining the asymptotic RG logaritms in both cases. From
inspection of these Figures a number of conclusions can be drawn. They
are listed in the final Section V.

\section{Conclusions}

In this paper we have extended to the case of the process
$e^+e^-\to t\bar t$ the study of the high energy behaviour 
of four-fermion processes that we had undertaken in previous works for
the case of light fermions. First we have shown how the $Z$-peak
subtracted representation can still be used to describe this process,
by taking as inputs the measurements of the charm-anticharm process
at the $Z$-peak, and by putting inside the subtracted 
functions the difference
between the $t\bar t$ and the $c\bar c$ one loop effects. 
We have then applied
this method in order to obtain well-defined predictions for the
high energy behaviour of the various observables that can be 
experimentally studied in $e^+e^-\to t\bar t$. We have made
illustrations with the one loop effects that appear in the SM and in
the MSSM. The results of this investigation are now summarized.

1) The leading electroweak effect at the one loop level is quite
sizeable in the TeV region in all observables, with the only (expected)
exception of the forward-backward asymmetry, where the squared Sudakov
logarithms are practically vanishing, as a consequence of 
a general rule already
discussed in Ref.\cite{slog}. The effect increases with energy,
following a trend that is drastically different from that of the smooth
and much smaller pure RG approximation, and it appears therefore to be
essentially governed by the various logarithms of "Sudakov-type".\\

2) The leading effects for top production are systematically larger
than those in the corresponding lepton or "light" $(u,d,s,c,b)$ quark
production observables. This is valid both in the SM and in the MSSM
situation. In the latter case, top production exhibits also in the
leading terms a drastic dependence on $\tan\beta$, much stronger than
that of bottom production (shown in Ref.\cite{slog}).\\

3) The validity of a one loop perturbative expansion for top production
seems to us neither too likely nor too unlikely in the TeV regime.
Around one TeV, all the effects are substancially under control (e.g.
below the ten percent level, assumed to  be a rough threshold for the
reliability of the approximation). In the CLIC region (3-5 TeV) the ten
percent boundary is systematically crossed, and the effect in
the MSSM can reach values varying from fifteen 
to twentyfive percent in the cross
section, depending on $\tan\beta$ (the SM effect is smaller in this
case; an opposite situation characterizes the two polarization
asymmetries). For the specific purposes of a very high precision test,
this would strongly motivate a (hard) two loop calculation of the most
relevant (in practice, the logarithms of "Sudakov-type") effects
\cite{XXX}. On the other hand, one could argue that possible neglected
terms, e.g. constant ones, might somehow reduce the size of the effect.
Our personal feeling, motivated by the previous experience for light
fermion production \cite{log}, \cite{mt}, 
is that in the SM case these extra
terms can reduce the effect, but not drastically (i.e. at the few
percent reduction level). In the MSSM case, this feeling remains to be
investigated in some more detail, although it is confirmed by a partial
previous analysis performed in Ref.\cite{slog}. Certainly, if one moves
to the 10 TeV region, where the leading logarithms should provide a
rather reliable approximation, the forty percent relative effect in
the top cross section shown in Fig.3 appears hardly compatible with a
one-loop truncation of the electroweak perturbative expansion.\\

4) The strong dependence on $\tan\beta$ of the leading asymptotic terms
appears to be a special characteristic of the top production in the TeV
regime. This is a consequence of the "massive" linear logarithms of
"Sudakov-type", proportional to $m^2_t$ (and also in the SUSY case, to
$m^2_b$) and generated by the final top vertex. To visualize the
numerical dependence, we have plotted in the final Fig.11-13 the
variation of the leading effects with $\tan\beta$, at the CLIC
reference point" $\sqrt{q^2}=3~TeV$. As one sees, the numerical
dependence has some features that appear to us potentially interesting.
Assuming a typical "visibility parameter" of one percent, that
corresponds to an integrated luminosity of about $10^3~fb^{-1}$, we
notice that:

a) The effect should be largely visible (e.g. it is
around the 10 percent level when $tan\beta$ varies from $1$ to $10$)
in the case of the cross section $\sigma_t$; it remains less strongly
but still potentially visible in
the polarized asymmetries $A^{LR}_t$ and $A_t$ (around the 
few percent level varying $tan\beta$ from $1$ to $10$); it is 
irrelevant (10 times smaller) in $A_{t,FB}$.\\

b) It varies from $-14$ to $-5$ and from $-5$ to $-9$ percent in
$\sigma_t$ and from $+4$ to $+1$ and from $+1$ to $-1$ percent 
in the polarized
asymmetries $A^{LR}_t$ and $A_t$ 
when $\tan\beta$ varies from $1$ to $10$ and from $10$ to $40$. 
Therefore, it is
in principle sensitive to the large $tan\beta$ region.\\

These features, if retained by a more complete approximation e.g. that
includes possible constant terms, would make top production in the CLIC
regime a promising and, in a certain sense, unique "$\tan\beta$
detector". A dedicated analysis with this specific purpose is already
being actively performed \cite{ttform}.\\

{\bf Acknowledgments}: This work has been partially
supported by the European
Community grant  HPRN-CT-2000-00149.

\newpage

\appendix

\section{Asymptotic logarithmic contributions in the MSSM}

\subsection{SM contributions}

In order to allow an easy comparison of the above SUSY contributions
with the SM ones we now recall, in the next three subsections,
 the results obtained in \cite{log},
\cite{mt} for the same four gauge invariant functions.\\

{\bf (a) Universal SM contributions}

\bq
\tilde{\Delta}^{(RG)}_{\alpha}(q^2,\theta)\to
{\alpha(\mu^2)\over12\pi}\left({32\over3}N-21\right)
\ln({q^2\over \mu^2})
\label{daRG}\eq

\bq
R^{(RG)}(q^2,\theta)\to-{\alpha(\mu^2)\over4\pi
s^2_Wc^2_W}\left({20-40c^2_W+32c^4_W\over9}N+{1-2c^2_W-42c^4_W\over6}
\right)
\ln({q^2\over \mu^2})
\label{RRG}\eq

\bq
V^{(RG)}_{\gamma Z}(q^2,\theta)
=V^{(RG)}_{Z\gamma}(q^2,\theta)
\to{\alpha(\mu^2)\over3\pi
s_Wc_W}\left({10-16c^2_W\over6}N+{1+42c^2_W\over8}\right)
\ln({q^2\over \mu^2})
\label{VRG}\eq

{\bf(b) $m_{t,b}$-independent terms in 
SM non-universal contributions to $l^+l^-\to t\bar t$
(same as in $u\bar u,~ c\bar c$)}

\bqa
\tilde{\Delta}^{(S)}_{\alpha,lf}(q^2,\theta)&\to&
{5\alpha\over4\pi}\ln{q^2\over M^2_W}
+{\alpha\over12\pi}\ln^2{q^2\over M^2_W}
+{\alpha(2-v^2_l-v^2_t)\over64\pi s^2_Wc^2_W}\left(3\ln{q^2\over M^2_Z}
-\ln^2{q^2\over M^2_Z}\right)\nonumber\\
&&-{\alpha\over2\pi}\left(\ln^2{q^2\over M^2_W}+2\ln{q^2\over
M^2_W}\ln{1+cos\theta\over2}\right)\nonumber\\
&&-{\alpha\over256\pi Q_f
s^4_Wc^4_W}(1-v^2_l)(1-v^2_t)\ln{q^2\over
M^2_Z}\ln{1+cos\theta\over1-cos\theta}\nonumber\\&&
\eqa

\bqa
R^{(S)}_{lf}(q^2,\theta)&\to&
-{3\alpha\over4\pi s^2_W}\left(1-{5s^2_W\over3}\right)
\ln{q^2\over M^2_W}
-{\alpha\over4\pi s^2_W}
\left(1-{s^2_W\over3}\right)
\ln^2{q^2\over M^2_W}\nonumber
\\
&& 
-{\alpha(2+3v^2_l+3v^2_t)\over64\pi s^2_Wc^2_W}\left(3\ln{q^2\over M^2_Z}
-\ln^2{q^2\over M^2_Z}\right)
+{\alpha c^2_W\over2\pi s^2_W}\left(\ln^2{q^2\over M^2_W}+2\ln{q^2\over
M^2_W}\ln{1+cos\theta\over2}\right)\nonumber\\
&&
+{\alpha\over4\pi s^2_Wc^2_W}v_lv_t\ln{q^2\over
M^2_Z}\ln{1+cos\theta\over1-cos\theta}\nonumber\\&&
\eqa

\bqa
V^{(S)}_{\gamma Z,lf}(q^2,\theta)&\to&
{\alpha \over8\pi c_Ws_W}\left[(3-10c^2_W)
\ln{q^2\over M^2_W}-\left(1+{2\over3}c^2_W\right)
\ln^2{q^2\over M^2_W}\right]\nonumber\\
&&
-\left[{\alpha v_l(1-v^2_l)\over128\pi s^3_Wc^3_W}+
{\alpha|Q_t|v_t\over8\pi s_Wc_W}\right]
\left(3\ln{q^2\over M^2_Z}-\ln^2{q^2\over M^2_Z}\right)\nonumber\\
&&
+{\alpha c_W\over2\pi s_W}
\left(\ln^2{q^2\over M^2_W}+2\ln{q^2\over
M^2_W}\ln{1+cos\theta\over2}\right)\nonumber\\
&&
+{\alpha\over32\pi s^3_Wc^3_W}v_t(1-v^2_l)\ln{q^2\over
M^2_Z}\ln{1+cos\theta\over1-cos\theta}\nonumber\\&&
\eqa

\bqa
V^{(S)}_{Z\gamma,lf}(q^2,\theta)&\to&
{\alpha \over8\pi cs}\left[(10s^2_W-9)
\ln{q^2\over M^2_W}-\left(1-{2\over3}s^2_W\right)
\ln^2{q^2\over M^2_W}\right]\nonumber\\
&&
-\left[{\alpha v_t(1-v^2_t)\over128\pi |Q_t|s^3_Wc^3_W}
+{\alpha v_l\over8\pi s_Wc_W}\right]
\left(3\ln{q^2\over M^2_Z}
-\ln^2{q^2\over M^2_Z}\right)\nonumber\\
&&
+{\alpha c_W\over2\pi s_W}
\left(\ln^2{q^2\over M^2_W}+2\ln{q^2\over
M^2_W}\ln{1+cos\theta\over2}\right)\nonumber\\
&&
+{\alpha\over32\pi Q_t s^3_Wc^3_W}v_l(1-v^2_t)\ln{q^2\over
M^2_Z}\ln{1+cos\theta\over1-cos\theta}\nonumber\\&&
\eqa

\noindent
where 
$v_l=1-4 \,s^2_W,\;\;v_t=1-4 \,|Q_t|\,s^2_W $.\par
In each of the above equations, we have successively added the
contributions coming from triangles containing one or two $W$,
from triangles containing one $Z$, from $WW$ box and finally from $ZZ$
box.\\

{\bf (c) $m_{t,b}$-dependent terms in non-universal
SM contributions to $l^+l^-\to t\bar t$}

\bqa
\tilde{\Delta}_{\alpha,lt}(q^2)&\to&
\tilde{\Delta}_{\alpha,lc}(q^2)-
\frac{\alpha}{24\pi s_W^2} \ln q^2
\left[
(3-2s^2_W)
\frac{m_t^2}{M_W^2}+2s_W^2\frac{m_b^2}{M_W^2}
\right]
\eqa

\bqa
R_{lt}(q^2)&\to&
R_{lc}(q^2)+
\frac{\alpha}{16\pi s^2_W} \ln q^2
\left[
 \left(1+\frac{4s_W^2}{3}\right) \frac{m_t^2}{M_W^2}
+\left(1-\frac{4s^2_W}{3}\right) \frac{m^2_b}{M^2_W}
\right]
\eqa

\bqa
V_{\gamma Z, lt}(q^2)&\to&
V_{\gamma Z, lc}(q^2)
-{\alpha c_W\over12 \pi s_W} \ln q^2
\left(
\frac{m^2_t}{M^2_W}-\frac{m_b^2}{M_W^2}
\right)
\eqa

\bqa
V_{Z\gamma , lt}(q^2)&\to&
V_{Z\gamma , lc}(q^2)
-{\alpha\over 16\pi s_W c_W} \ln q^2
\left(1-\frac{4s^2_W}{3}\right)
\left(
\frac{m^2_t}{M^2_W}-\frac{m_b^2}{M_W^2}
\right)
\eqa\\

\subsection{Additional SUSY contributions}

{\bf (a)Universal ($\gamma,Z$-self-energy) SUSY contributions}\\

They arise from the bubbles (and associated tadpole diagrams) involving
internal L- and R- sleptons and squarks, charginos, neutralinos, 
as well as the charged and neutral Higgses and Goldstones
(subtracting the standard Higgs contribution):\\

\bqa
\tilde{\Delta}^{Univ}_{\alpha}(q^2)&\to&
{\alpha\over4\pi}\left(3+{16N\over9}\right)\ \ln q^2
\label{DAun}\eqa

\bqa
R^{Univ}(q^2)&\to&
-{\alpha\over4\pi s^2_W c^2_W}
\left[{13-26s^2_W+18s^4_W\over6}
+(3-6s^2_W+8s^4_W){2N\over9}\right]\ \ln q^2\nonumber\\
\label{Run}\eqa

\bqa
V^{Univ}_{\gamma Z}(q^2)=V^{Univ}_{Z\gamma}(q^2)&\to&
-{\alpha\over4\pi s_W c_W}
\left[{13-18s^2_W\over6}+(3-8s^2_W){2N\over9}\right]\ \ln q^2
\label{Vun}\eqa

\noindent
where N is the number of slepton and squark families. These terms
contribute to the RG effects.\\

We then consider the non-universal SUSY contributions.
These are the contributions coming from triangle diagrams connected
either to the initial $l^+l^-$ or to the final $t\bar t$ lines, and
containing SUSY partners, sfermions $\tilde{f}$, charginos or
neutralinos $\chi_i$, or
SUSY Higgses (see Fig.2); external fermion self-energy diagrams
are added making the total contribution finite. These
non universal terms
consist in $m_{t,b}$-independent terms  and in $m_{t,b}$-dependent terms
(quadratic $m^2_t$ and $m^2_b$ terms) given below:\\

{\bf (b) $m_{t,b}$-independent terms in SUSY
contributions to $l^+l^-\to t\bar t$
(same as for $u\bar u,~c\bar c$)}

\bqa
\tilde{\Delta}_{\alpha,lt}(q^2)&\to&
{\alpha\over\pi}\ \ln q^2\
{-71+82s^2_W\over 72c^2_W}
\label{DAnunu}\eqa

\bqa
R_{lt}(q^2)&\to&
{\alpha\over\pi}\ \ln q^2\
{27-67s^2_W+82s^4_W\over 72s^2_Wc^2_W}
\label{Rgnunu}\eqa

\bqa
V_{\gamma Z, lt}(q^2)&\to&{\alpha\over\pi}\ \ln q^2
\ {63-200s^2_W+164s^4_W\over 144s_Wc^3_W}
\label{Vgznunu}\eqa

\bqa
V_{Z\gamma , lt}(q^2)&\to&{\alpha\over\pi}\ \ln q^2\
{81-240s^2_W+164s^4_W\over 144s_Wc^3_W}
\label{Vzgnunu}\eqa\\

{\bf (c) $m_{t,b}$-dependent terms in 
SUSY contributions to $l^+l^-\to t\bar t$}

\bqa
\tilde{\Delta}_{\alpha,lt}(q^2)&\to&
\tilde{\Delta}_{\alpha,lc}(q^2)-
\frac{\alpha}{24\pi s_W^2} \ln q^2
\left[
(3-2s^2_W)(1+2\cot^2\beta)
\frac{m_t^2}{M_W^2}+2s_W^2\frac{m_b^2}{M_W^2}(1+2\tan^2\beta)
\right]
\eqa

\bqa
R_{lt}(q^2)&\to&
R_{lc}(q^2)+
\frac{\alpha}{16\pi s^2_W} \ln q^2
\left[
 \left(1+\frac{4s_W^2}{3}\right)(1+2\cot^2\beta) \frac{m_t^2}{M_W^2}
+\left(1-\frac{4s^2_W}{3}\right) \frac{m^2_b}{M^2_W}(1+2\tan^2\beta)
\right]
\eqa

\bqa
V_{\gamma Z, lt}(q^2)&\to&
V_{\gamma Z, lc}(q^2)
-{\alpha c_W\over12 \pi s_W} \ln q^2
\left(
\frac{m^2_t}{M^2_W}(1+2\cot^2\beta)-\frac{m_b^2}{M_W^2}(1+2\tan^2\beta)
\right)
\eqa

\bqa
V_{Z\gamma , lt}(q^2)&\to&
V_{Z\gamma , lc}(q^2)
-{\alpha\over 16\pi s_W c_W} \ln q^2
\left(1-\frac{4s^2_W}{3}\right)
\left(
\frac{m^2_t}{M^2_W}(1+2\cot^2\beta)-\frac{m_b^2}{M_W^2}(1+2\tan^2\beta)
\right)
\eqa\\

\subsection{Non-universal massive MSSM contribution}

\vspace{0.5cm}

Finally we find interesting to sum up all the massive $m^2_t$ and
$m^2_b$ terms appearing in the MSSM (SM and SUSY non-universal massive
contributions to $l^+l^-\to t\bar t$). We remark that the net effect as
compared to the SM result is
a factor $2(1+\cot^2\beta)$ for the $m^2_t$ term and a factor 
$2(1+\tan^2\beta)$ for the $m^2_b$ one, a rule similar to the one
observed in the $b\bar b$ case \cite{slog}:

\bqa
\tilde{\Delta}_{\alpha,lt}(q^2)&\to&
\tilde{\Delta}_{\alpha,lc}(q^2)-
\frac{\alpha}{12\pi s_W^2} \ln q^2
\left[
(3-2s^2_W)(1+\cot^2\beta)
\frac{m_t^2}{M_W^2}+2s_W^2\frac{m_b^2}{M_W^2}(1+\tan^2\beta)
\right]
\eqa

\bqa
R_{lt}(q^2)&\to&
R_{lc}(q^2)+
\frac{\alpha}{8\pi s^2_W} \ln q^2
\left[
 \left(1+\frac{4s_W^2}{3}\right)(1+\cot^2\beta) \frac{m_t^2}{M_W^2}
+\left(1-\frac{4s^2_W}{3}\right) \frac{m^2_b}{M^2_W}(1+\tan^2\beta)
\right]
\eqa

\bqa
V_{\gamma Z, lt}(q^2)&\to&
V_{\gamma Z, lc}(q^2)
-{\alpha c_W\over6 \pi s_W} \ln q^2
\left(
\frac{m^2_t}{M^2_W}(1+\cot^2\beta)-\frac{m_b^2}{M_W^2}(1+\tan^2\beta)
\right)
\eqa

\bqa
V_{Z\gamma , lt}(q^2)&\to&
V_{Z\gamma , lc}(q^2)
-{\alpha\over 8\pi s_W c_W} \ln q^2
\left(1-\frac{4s^2_W}{3}\right)
\left(
\frac{m^2_t}{M^2_W}(1+\cot^2\beta)-\frac{m_b^2}{M_W^2}(1+\tan^2\beta)
\right)
\eqa\\

\newpage

\section{Helicity amplitudes and top spin density matrix}

A generic invariant amplitude

\bqa
&&A={e^2\over q^2}\bar u(t)[\gamma^{\mu}g^{\gamma}_{Vt}
+{d^{\gamma}\over m_t}(p-p')_{\mu}]v(\bar t)~.~
\bar v(l)\gamma_{\mu}g^{\gamma}_{Vl}u(l)+\nonumber\\
&&{e^2\over 4s^2_Wc^2_W(q^2-m^2_Z)}
\bar u(t)[\gamma^{\mu}(g^{Z}_{Vt}-
g^{Z}_{At}\gamma^5)+{d^{Z}\over m_t}(p-p')^{\mu}]v(\bar t)~.~
\bar v(l)\gamma_{\mu}(g^{Z}_{Vl}-g^{Z}_{Al}\gamma^5)u(l)
\label{Aeff}\eqa
\noindent
leads to the helicity amplitude

\bqa
F(\lambda_l,\lambda_t,\lambda_{\bar t})&=&
(2\lambda_l)e^2\sqrt{q^2}.\nonumber\\
&&.\{~{g^{\gamma}_{Vl}\over q^2}[~g^{\gamma}_{Vt}(
2m_tsin\theta\delta_{\lambda_t,\lambda_{\bar t}}-\sqrt{q^2}
(\lambda_t-\lambda_{\bar t})cos\theta-2\lambda_l\sqrt{q^2}
\delta_{\lambda_t,-\lambda_{\bar t}})
-{d^{\gamma}\over m_t}\beta^2_tq^2sin\theta
\delta_{\lambda_t,\lambda_{\bar t}}]
\nonumber\\
&&
+{g^{Z}_{Vl}-2\lambda g^{Z}_{Al}\over4s^2_Wc^2_W(q^2-m^2_Z)}[
g^{Z}_{Vt}(
2m_tsin\theta\delta_{\lambda_t,\lambda_{\bar t}}-\sqrt{q^2}
(\lambda_t-\lambda_{\bar t})cos\theta-2\lambda\sqrt{q^2}
\delta_{\lambda_t,-\lambda_{\bar t}})\nonumber\\
&&+
g^{Z}_{At}\beta_t\sqrt{q^2}(
cos\theta\delta_{\lambda_t,-\lambda_{\bar t}}
+2\lambda_l(\lambda_t-\lambda_{\bar t})
-{d^{Z}\over
m_t}\beta^2_tq^2sin\theta\delta_{\lambda_t,\lambda_{\bar t}}]\}
\label{Fhel}\eqa
\noindent
where $\lambda_l\equiv\lambda_{l^-}\equiv -\lambda_{l^+}=\pm {1\over2}$,
$\lambda_t=\pm {1\over2}$, $\lambda_{\bar t}=\pm {1\over2}$ and the
normalization is such that the differential cross section is given by
\bqa
{d\sigma\over dcos\theta} ={\beta_tN_t\over64\pi
q^2}\{(1-PP')[\rho^U(+,+)+\rho^U(-,-)]+(P-P')
[\rho^{LR}(+,+)+\rho^{LR}(-,-)]\}
\label{dsigt}\eqa
\noindent
with $\beta_t=\sqrt{1-{4m^2_t\over q^2}}$. $N_t$ is the colour factor
$3$, times the QCD correction factor. The top quark spin density
matrix is defined as

\bqa
\rho^U(\lambda_t,\lambda'_t)\equiv
{1\over2}\rho^{L+R}(\lambda_t,\lambda'_t)
&=&{1\over2}\sum_{\lambda_{\bar t}}
[F(\lambda_l=-{1\over2},\lambda_t,\lambda_{\bar t})
F^*(\lambda_l=-{1\over2},\lambda'_t,\lambda_{\bar t})\nonumber\\
&&+F(\lambda_l=+{1\over2},\lambda_t,\lambda_{\bar t})
F^*(\lambda_l=+{1\over2},\lambda'_t,\lambda_{\bar t})]
\label{rhoU}\eqa
\noindent
for unpolarized $l^{\pm}$ beams, and

\bqa
\rho^{LR}(\lambda_t,\lambda'_t)\equiv
{1\over2}\rho^{L-R}(\lambda_t,\lambda'_t)
&=&{1\over2}\sum_{\lambda_{\bar t}}
[F(\lambda_l=-{1\over2},\lambda_t,\lambda_{\bar t})
F^*(\lambda_l=-{1\over2},\lambda'_t,\lambda_{\bar t})\nonumber\\
&&-F(\lambda_l=+{1\over2},\lambda_t,\lambda_{\bar t})
F^*(\lambda_l=+{1\over2},\lambda'_t,\lambda_{\bar t})]
\label{rhoLR}\eqa
\noindent
for longitudinally polarized $l^-$ and $l^+$ beams with degrees $P$
and $P'$, respectively. \\
{}From eq.(\ref{dsigt}) one constructs
the usual four observables $\sigma_t$, $A_{FB,t}$, 
$A_{LR,t}$ and $A_t$. But from the density matrices eq.(\ref{rhoU},
\ref{rhoLR}) one can also construct other observables related to
the final top quark polarization.

At large $q^2$ the top quark polarization is only described by its
helicity (the transverse polarization vanishes). So one defines 
the following observables:

\newpage

(a) with unpolarized $l^+l^-$ beams:

--- the averaged top helicity\\

\bq
H_t={\int~ [\rho^U(+,+)-\rho^U(-,-)]~dcos\theta\over
\int~ [\rho^U(+,+)+\rho^U(-,-)]~dcos\theta }
\eq

--- its forward-backward asymmetry

\bq
H_{t,FB}={\int_{F-B}~ [\rho^U(+,+)-\rho^U(-,-)]~dcos\theta\over
\int~ [\rho^U(+,+)+\rho^U(-,-)]~dcos\theta }
\eq\\

(b) and  the same two quantities for the Left-Right $l^-$ polarization
asymmetry

--- the averaged polarized top helicity\\

\bq
H^{LR}_t={\int~ [\rho^{LR}(+,+)-\rho^{LR}(-,-)]dcos\theta\over
\int~ [\rho^{U}(+,+)+\rho^{U}(-,-)]~dcos\theta }
\eq

--- and its forward-backward asymmetry

\bq
H^{LR}_{t,FB}={\int_{F-B}~ [\rho^{LR}(+,+)-\rho^{LR}(-,-)]
~dcos\theta\over
\int~ [\rho^U(+,+)+\rho^U(-,-)]~dcos\theta }
\eq

These quantities can be measured through the decay of the top
quark into $W+b$, see for example the discussion in ref.\cite{GR}.\\

In the asymptotic regime, the $Z$-peak subtraction method described
in Section II implies the following expressions for the effective
photon and $Z$ couplings to be used in eq.(\ref{Fhel}):

\bqa
g^{\gamma}_{Vl}\to
Q_e\left[1+{1\over2}\tilde{\Delta}_{\alpha,lt}(q^2,\theta)\right]
\eqa
\bqa
g^{\gamma}_{Vt}\to
Q_t\left[1+{1\over2}\tilde{\Delta}_{\alpha,lt}(q^2,\theta)\right]
\eqa

\bqa
{|e|\over2s_Wc_W}g^{Z}_{Vl}\to-~{\tilde{v}_l\over2}
\gamma^{{1\over2}}_l\left[1-{1\over2}\hat{R}_{lt}(q^2,\theta)
-{4s_Wc_W\over\tilde{v}_l}V^{\gamma Z}_{lt}\right]
\eqa

\bqa
{|e|\over2s_Wc_W}g^{Z}_{Al}\to-~{1\over2}
\gamma^{{1\over2}}_l\left[1-{1\over2}\hat{R}_{lt}(q^2,\theta)\right]
\eqa

\bqa
{|e|\over2s_Wc_W}g^{Z}_{Vt}\to {\tilde{v}_c\over2}
\gamma^{{1\over2}}_c\left[1-{1\over2}\hat{R}_{lt}(q^2,\theta)
-{8s_Wc_W\over3\tilde{v}_c}\hat{V}^{Z\gamma}_{lt}\right]
\eqa

\bqa
{|e|\over2s_Wc_W}g^{Z}_{At}\to{1\over2}
\gamma^{{1\over2}}_c\left[1-{1\over2}\hat{R}_{lt}(q^2,\theta)\right]
\eqa
\noindent
where $\tilde{v}_f=1-4|Q_f|s^2_{W,f}$, $s^2_{W,f}$ being the effective
angle measured at $Z$ peak in the channel $l^+l^-\to f\bar f$, and

\bqa
\gamma_f\equiv{48\pi\Gamma(Z\to f\bar f)
\over N_{Zf}M_Z(1+\tilde{v}^2_f)}
\eqa
\noindent
$N_{Zf}$ being the colour factor 3, times the QCD correction factor in
the $f\bar f$ channel at $Z$-peak.\\

Complete results including non asymptotic contributions and new Lorentz
structures will be given in the forthcoming paper \cite{ttform}.

\begin{figure}[p]
\vspace*{1cm}
\[
\epsfig{file=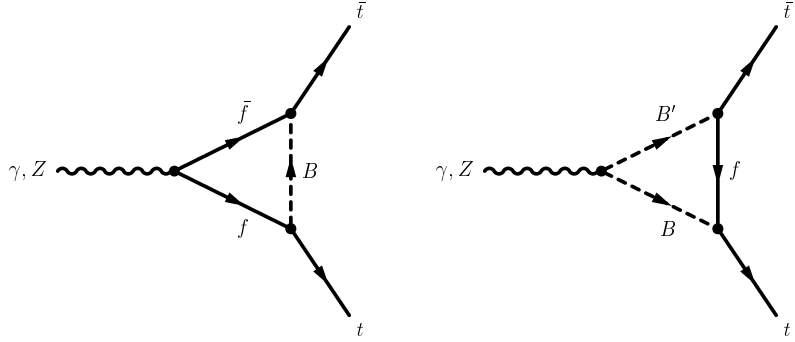,height=4.5cm}
\]
\vspace*{0.5cm}
\caption[1]{Triangle SM diagrams contributing to the asymptotic
logarithmic behaviour in the energy; $f$ represent $t$ or $b$ quarks,
$B$ represent $W^{\pm}$, $\Phi^{\pm}$ or $Z$, $G^{0}$, $H_{SM}$.}
\label{figsm}
\end{figure}\begin{figure}[p]

\[
\epsfig{file=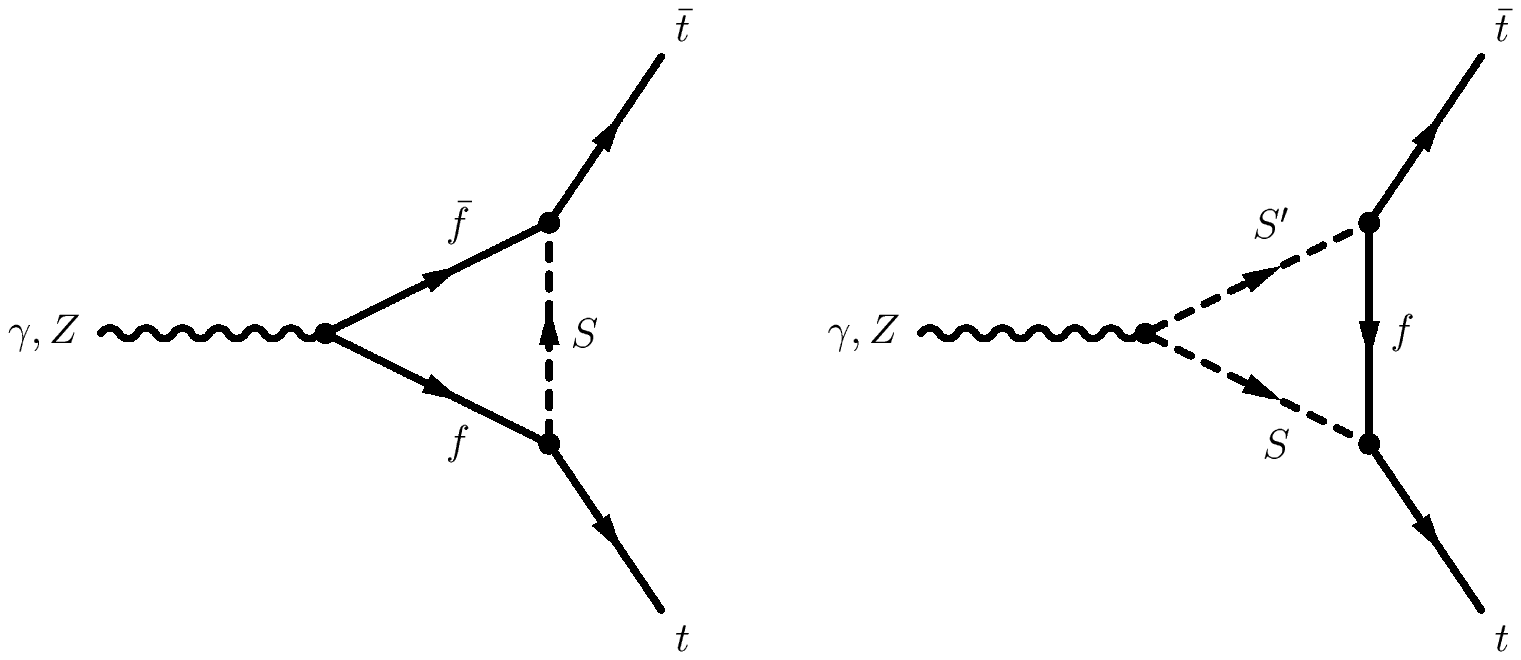,height=4.5cm}
\]
\vspace*{0.5cm}
\label{figS}
\end{figure}\begin{figure}[p]
\vspace*{-2cm}
\[
\epsfig{file=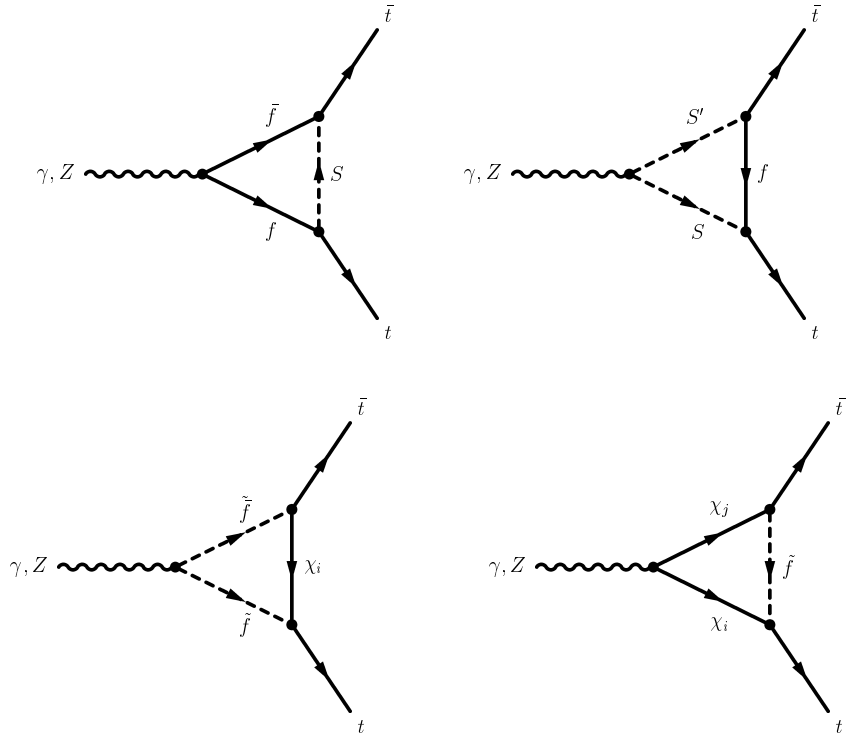,height=4.5cm}
\]
\caption[2]{Triangle diagrams with SUSY Higgs and with SUSY
partners contributing 
to the asymptotic logarithmic behaviour in the energy; 
$f$ represent $t$ or $b$ quarks; $S$ 
represent charged or neutral Higgs bosons $H^{\pm}$,
$A^0$, $H^0$, $h^0$ or Goldstone $G^0$; $\tilde f$ represent
stop or sbottom states; $\chi$ represent charginos or neutralinos.}
\label{figsusy}
\end{figure}
\newpage
\begin{figure}[p]
\vspace*{-2cm}
\[
\epsfig{file=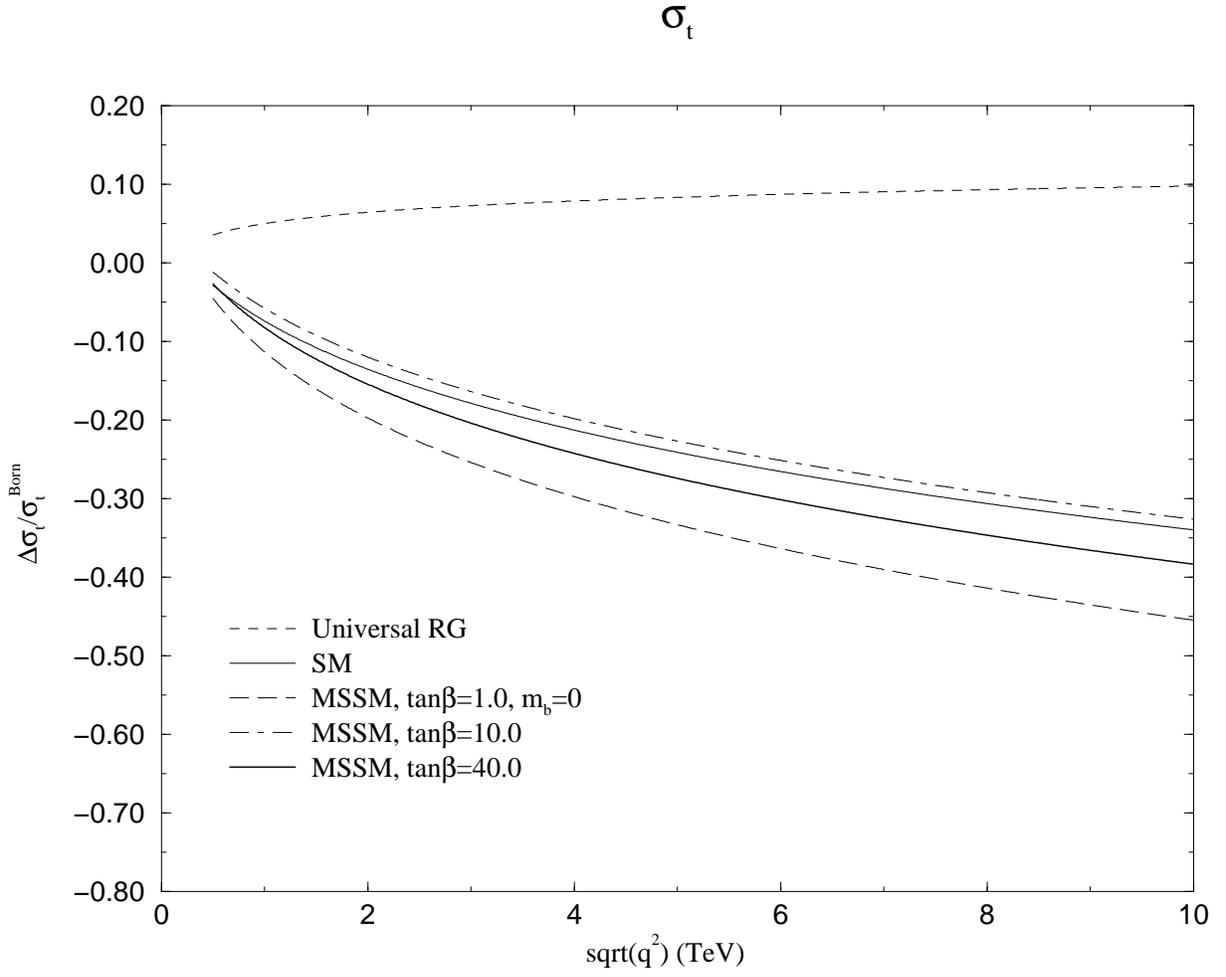 ,height=16cm,angle=-90}
\]
\caption[5]{Relative effects in $\sigma_t$ due to the asymptotic 
logarithmic terms.
The Born expression for large $q^2$ is 
$182\ \mbox{fb}/(q^2/{\rm TeV}^2)$.}
\label{sigt}
\end{figure}
\newpage
\begin{figure}[p]
\vspace*{-2cm}
\[
\epsfig{file=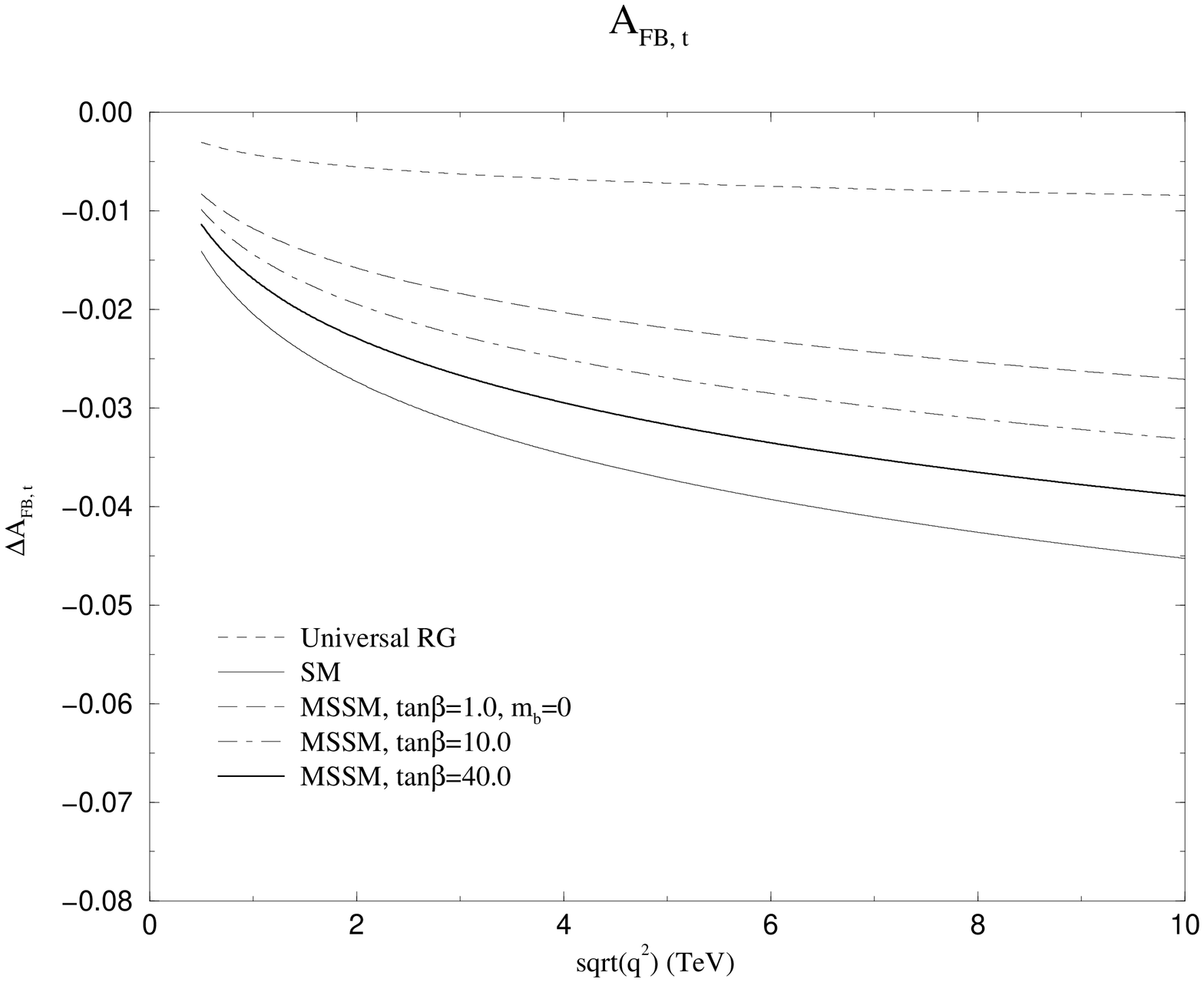 ,height=16cm,angle=-90}
\]
\caption[5]{Absolute effects in $A_{FB, t}$ due to the asymptotic 
logarithmic terms. The Born value for large $q^2$ is $0.607$.}
\label{afbt}
\end{figure}
\newpage
\begin{figure}[p]
\vspace*{-2cm}
\[
\epsfig{file=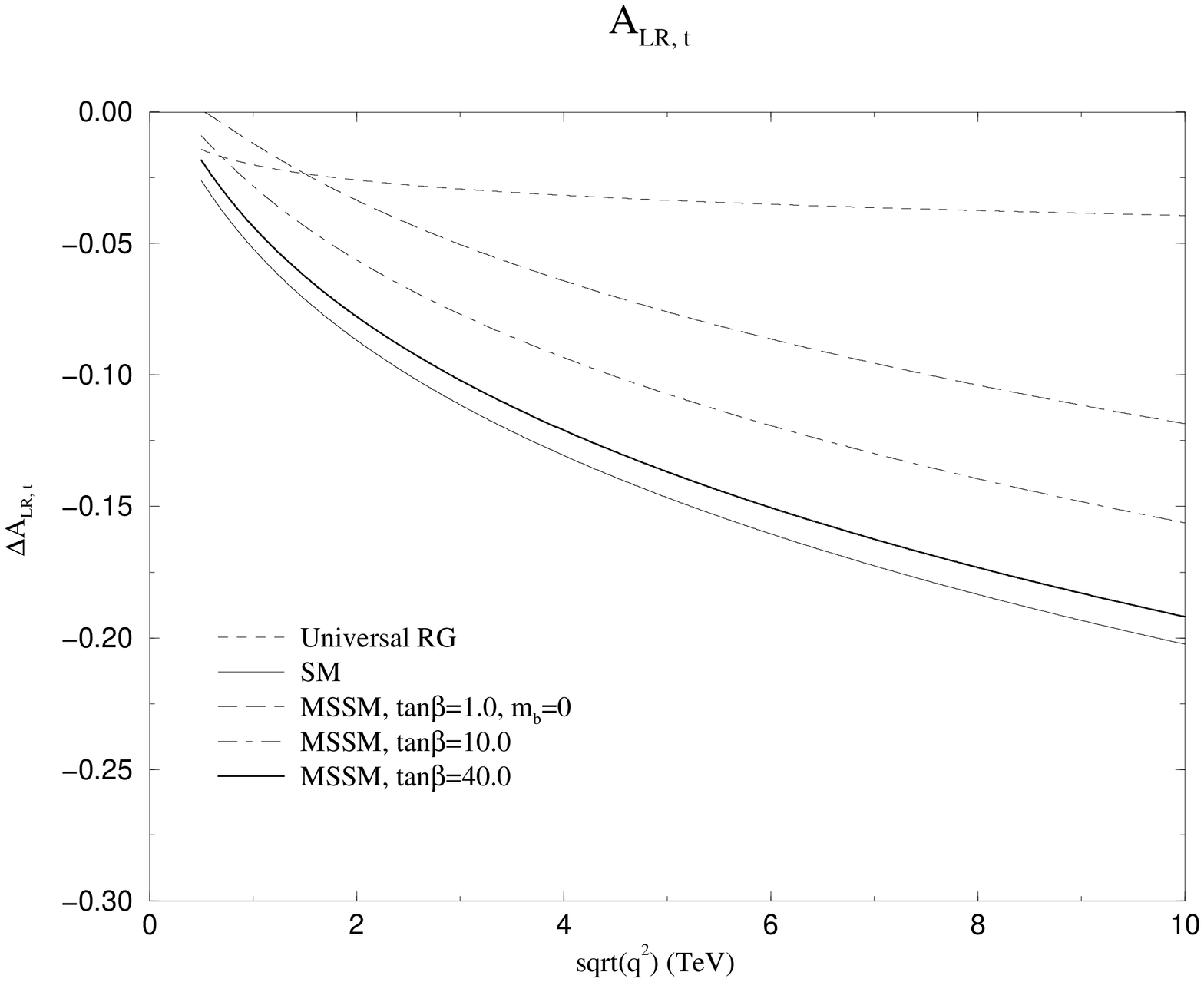 ,height=16cm,angle=-90}
\]
\caption[5]{Absolute effects in $A_{LR, t}$ due to the asymptotic 
logarithmic terms. The Born value for large $q^2$ is $0.336$.}
\label{alrt}
\end{figure}
\newpage
\begin{figure}[p]
\vspace*{-2cm}
\[
\epsfig{file=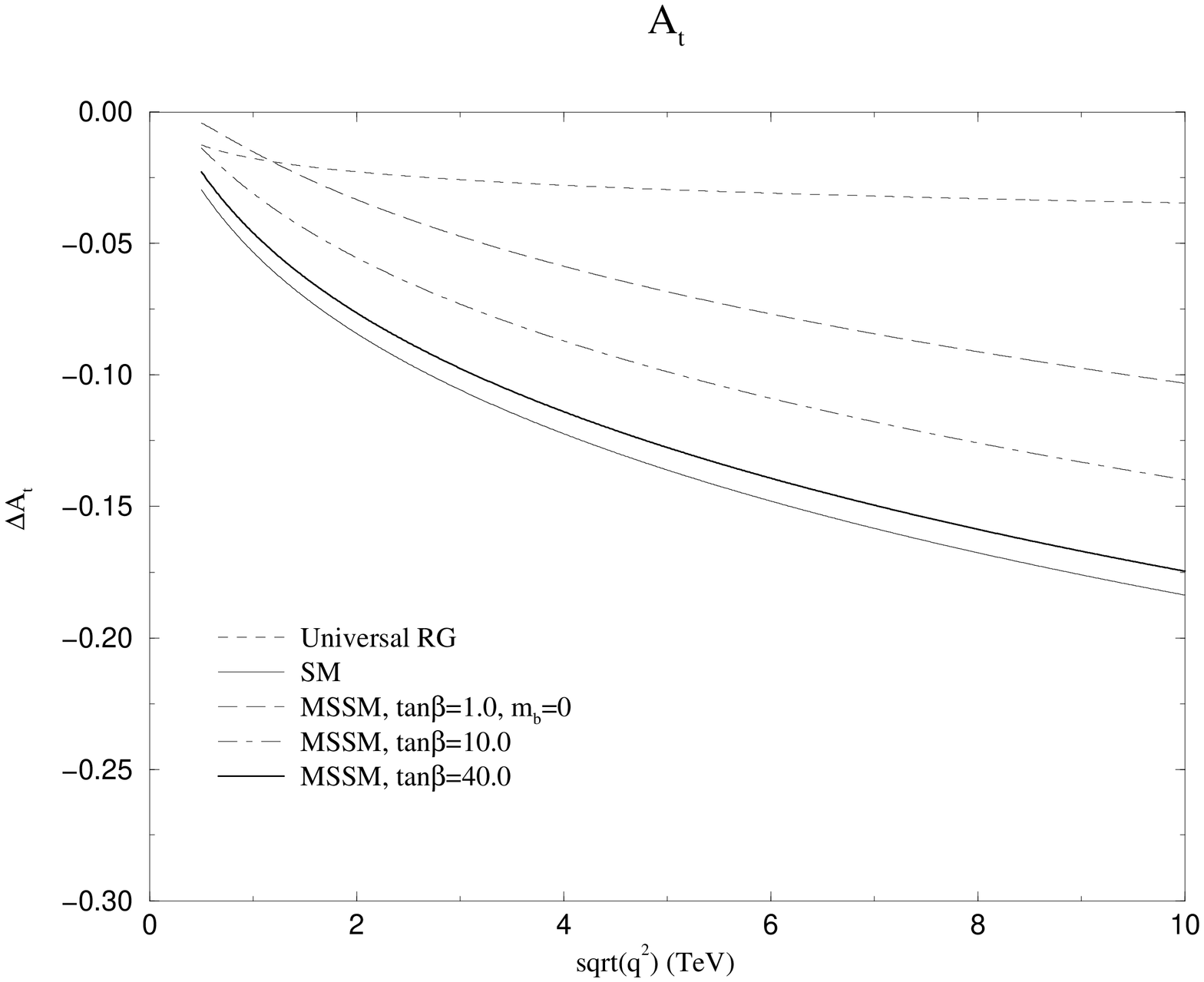  ,height=16cm,angle=-90}
\]
\caption[5]{Absolute effects in $A_{t}$ due to the asymptotic 
logarithmic terms. The Born value for large $q^2$ is $0.164$.}
\label{at}
\end{figure}
\newpage
\begin{figure}[p]
\vspace*{-2cm}
\[
\epsfig{file=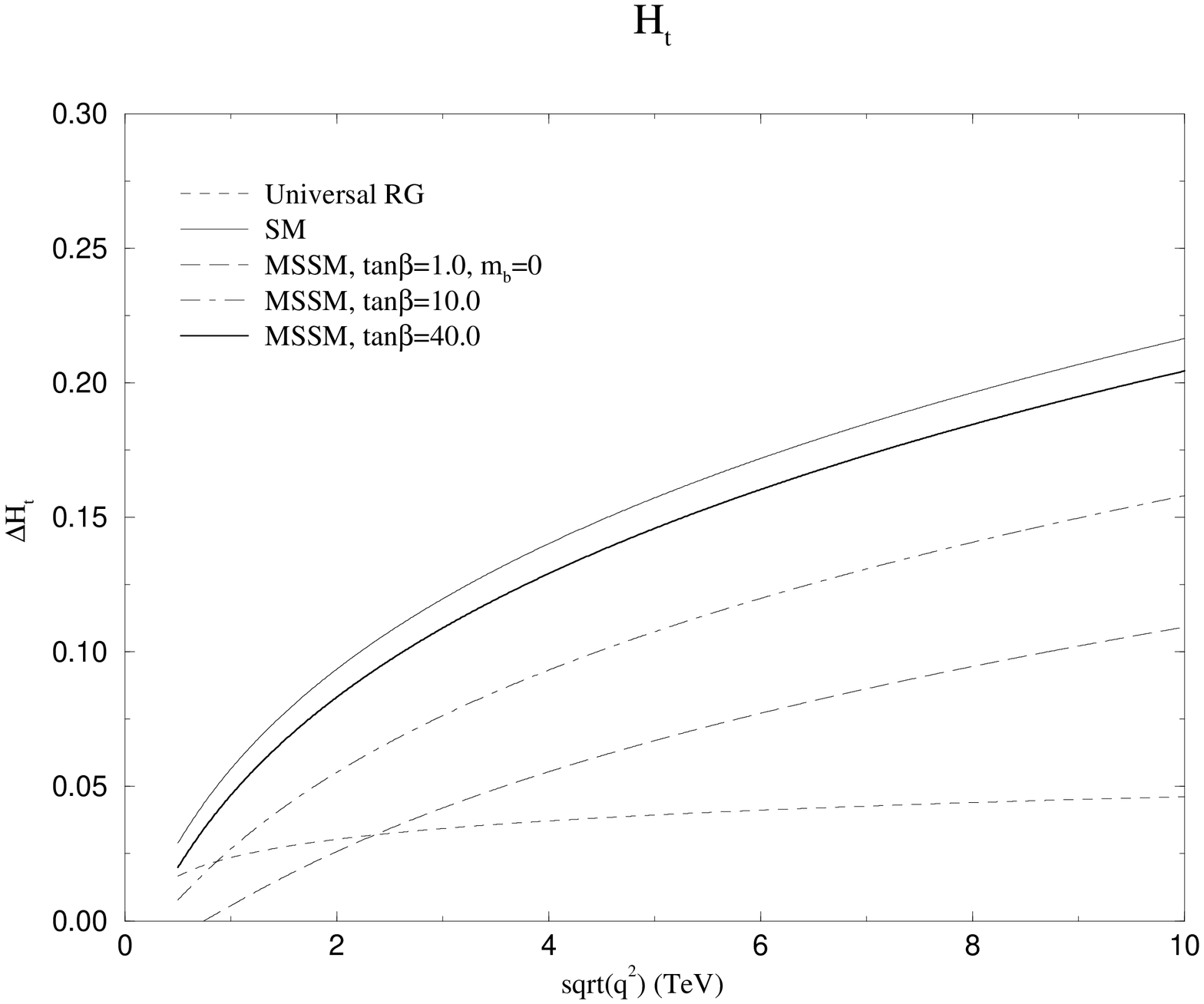  ,height=16cm,angle=-90}
\]
\caption[5]{Absolute effects in $H_{t}$ due to the asymptotic 
logarithmic terms. The Born value for large $q^2$ is $-0.219$.}
\label{ht}
\end{figure}
\newpage
\begin{figure}[p]
\vspace*{-2cm}
\[
\epsfig{file=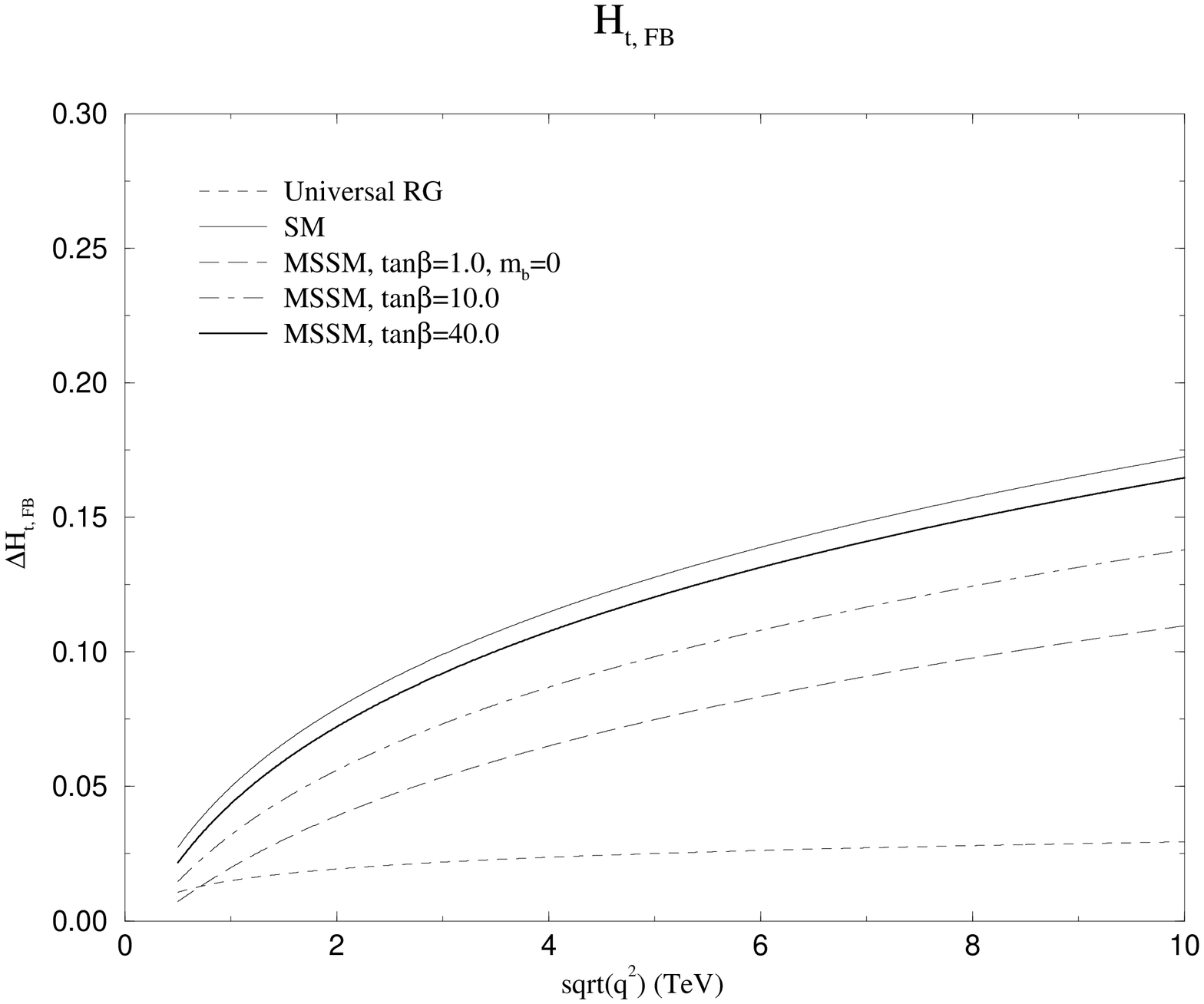  ,height=16cm,angle=-90}
\]
\caption[5]{Absolute effects in $H_{t,FB}$ due to the asymptotic 
logarithmic terms. The Born value for large $q^2$ is $-0.252$.}
\label{htfb}
\end{figure}
\newpage
\begin{figure}[p]
\vspace*{-2cm}
\[
\epsfig{file=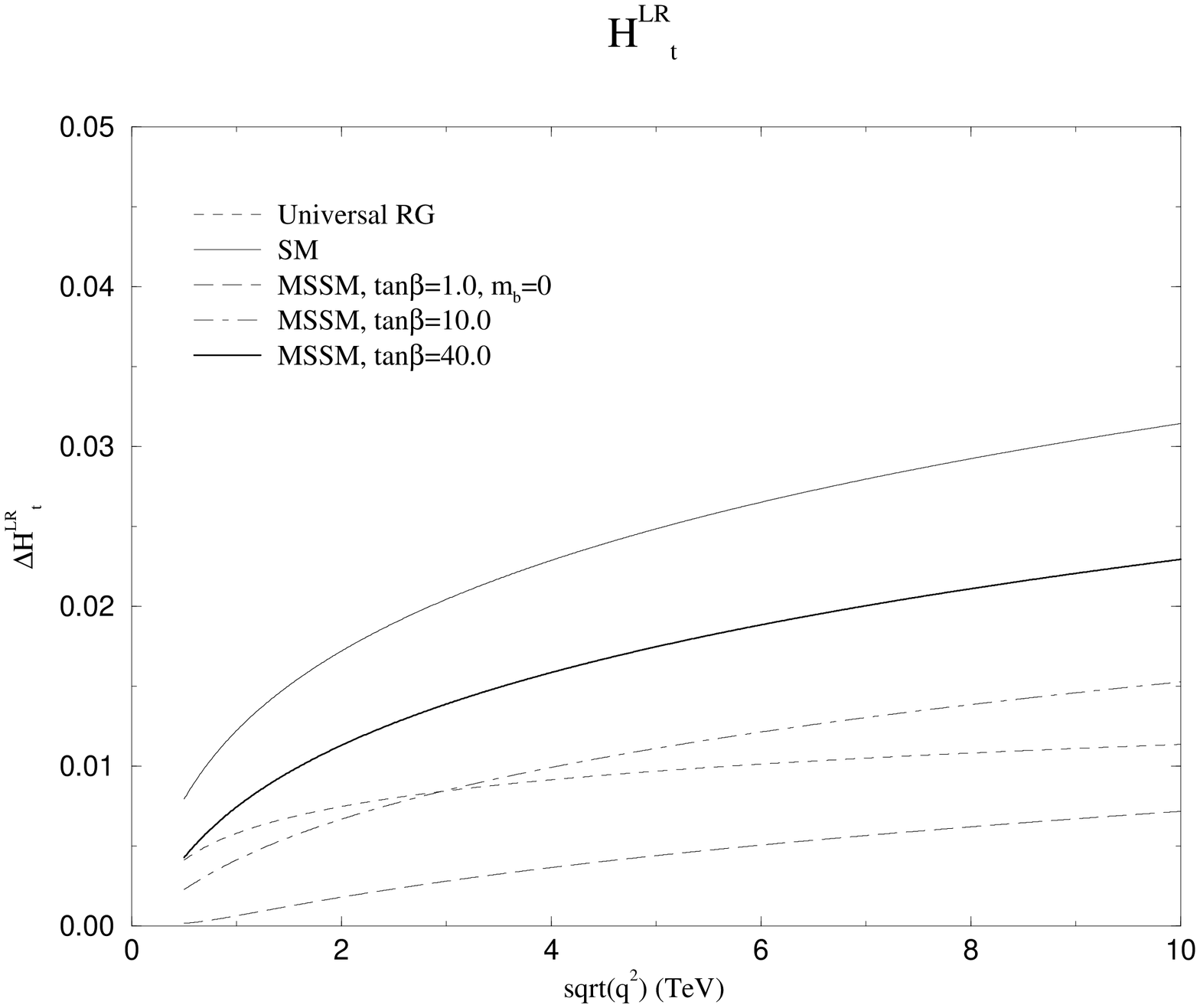  ,height=16cm,angle=-90}
\]
\caption[5]{Absolute effects in $H^{LR}_{t}$ due to the asymptotic 
logarithmic terms. The Born value for large $q^2$ is $-0.809$.}
\label{hlrt}
\end{figure}
\newpage
\begin{figure}[p]
\vspace*{-2cm}
\[
\epsfig{file=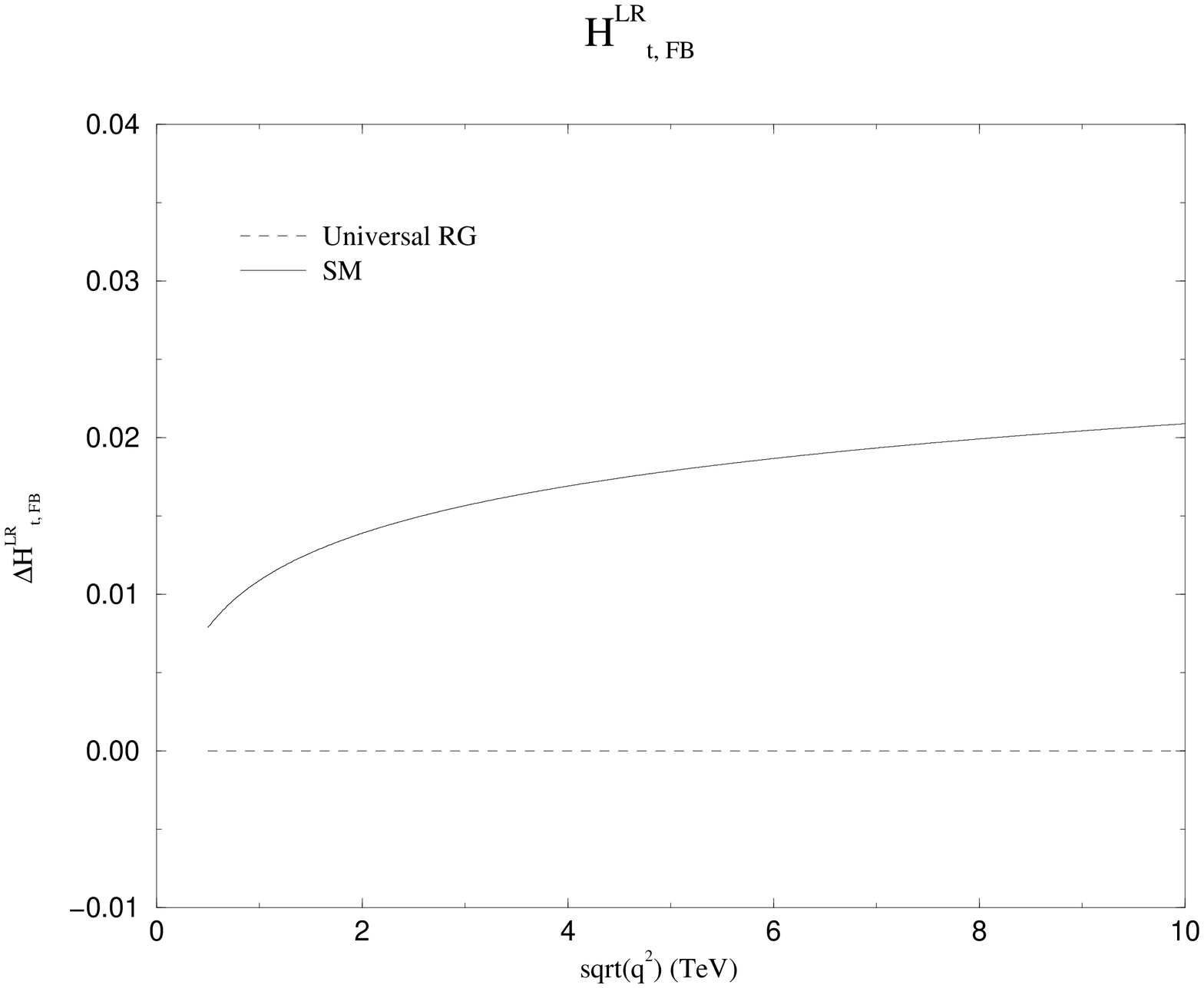   ,height=16cm,angle=-90}
\]
\caption[5]{Absolute effects in $H^{LR}_{t,FB}$ due to the asymptotic 
logarithmic terms (only SM box terms contribute). 
The Born value for large $q^2$ is $-0.750$.}
\label{hfbpolt}
\end{figure}
\newpage
\begin{figure}[p]
\vspace*{-2cm}
\[
\epsfig{file=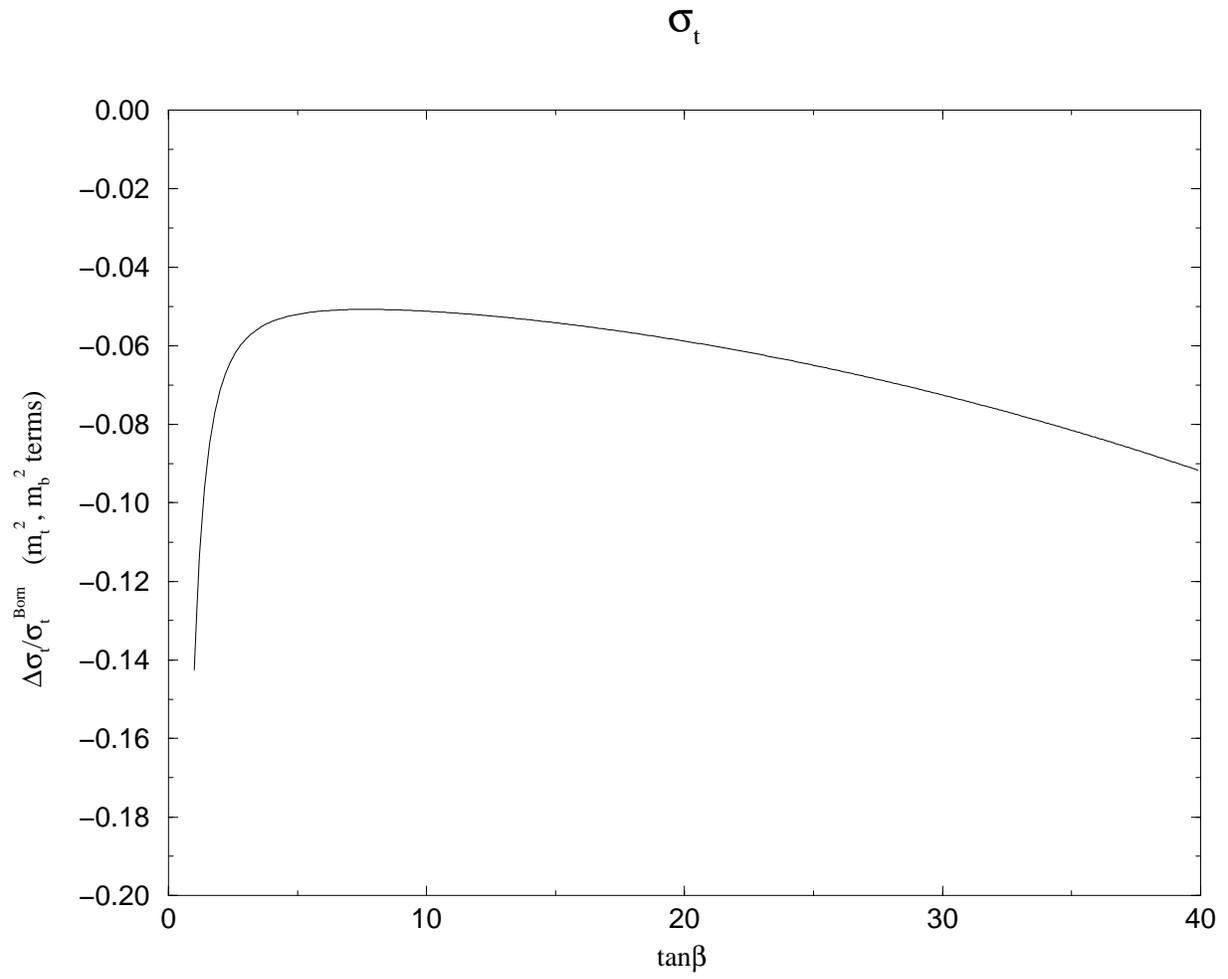   ,height=16cm,angle=-90}
\]
\caption[5]{Relative effects in $\sigma_t$ due to the asymptotic 
$m^2_t$ and $m^2_b$ logarithmic terms versus $\tan\beta$.}
\label{sigtbeta}
\end{figure}
\newpage
\begin{figure}[p]
\vspace*{-2cm}
\[
\epsfig{file=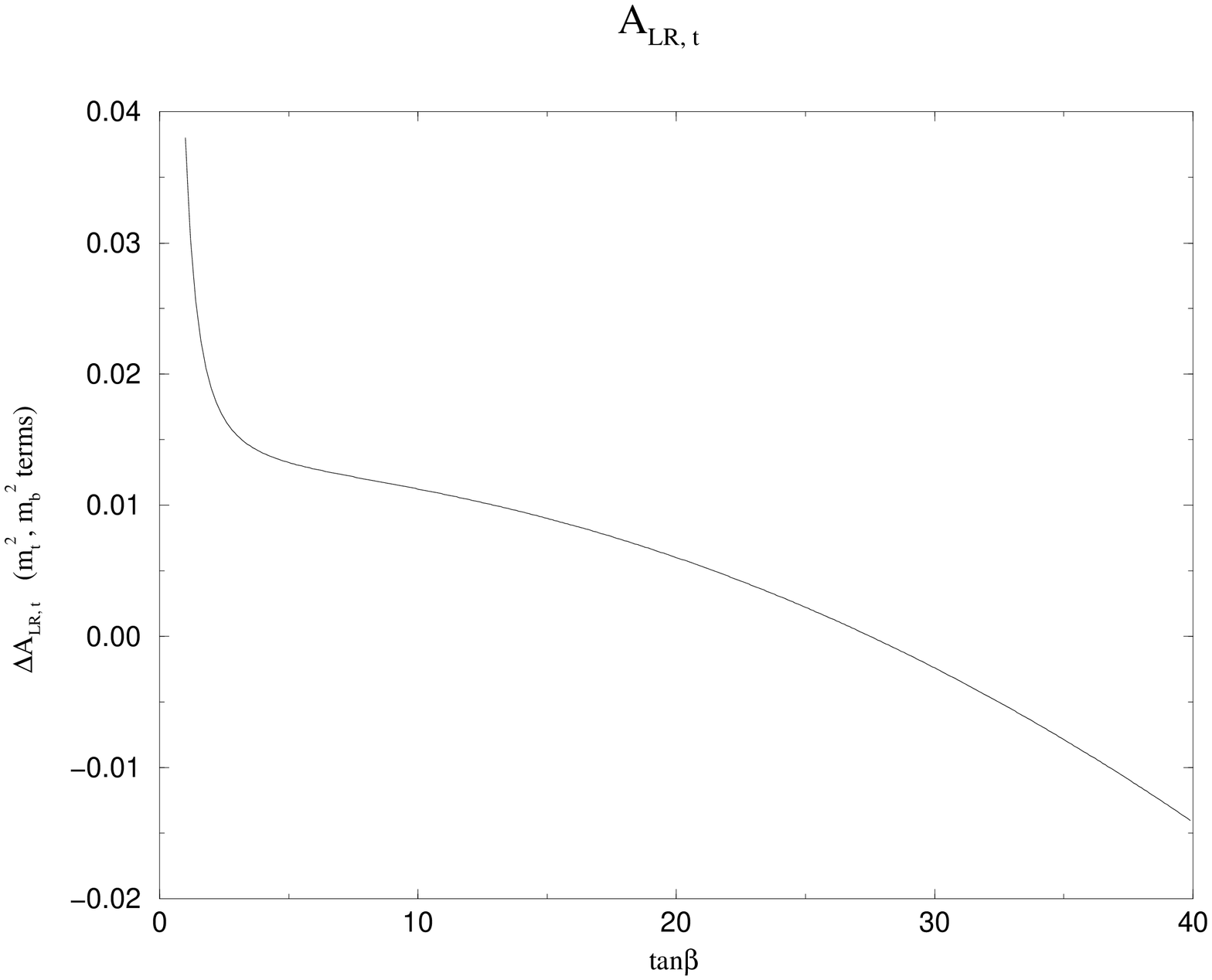   ,height=16cm,angle=-90}
\]
\caption[5]{Absolute effects in $A^{LR}_t$ due to the asymptotic 
$m^2_t$ and $m^2_b$ logarithmic terms versus $\tan\beta$.}
\label{alrtbeta}
\end{figure}
\newpage
\begin{figure}[p]
\vspace*{-2cm}
\[
\epsfig{file=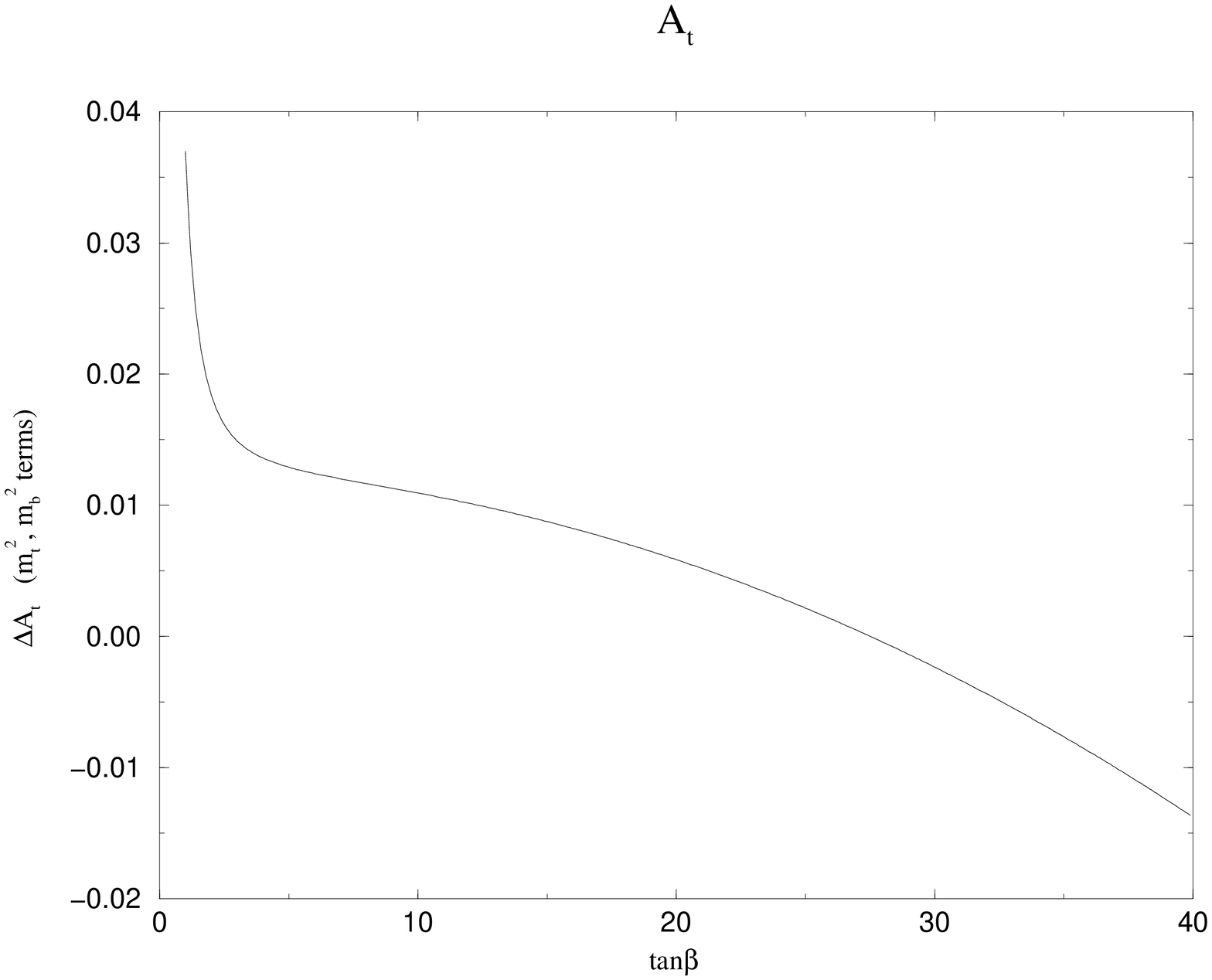   ,height=16cm,angle=-90}
\]
\caption[5]{Absolute effects in $A_{t}$ due to the asymptotic 
$m^2_t$ and $m^2_b$ logarithmic terms versus $\tan\beta$.}
\label{afbpoltbeta}
\end{figure}

\end{document}